\documentclass[10pt, twocolumn]{IEEEtran}

\usepackage{graphicx}
\usepackage{amsmath}
\usepackage[noend]{algpseudocode}
\usepackage{algorithmicx,algorithm}
\usepackage{amsthm}
\usepackage{amsfonts}
\usepackage{subfigure} 
\usepackage{color}
\usepackage{cite}
\usepackage{setspace}

\usepackage{amssymb}
\usepackage{tabularx}

\usepackage{xcolor}
\usepackage{xpatch}
\makeatletter
\def\changeBibColor#1{%
	\in@{#1}{}
	\ifin@\color{blue}\else\normalcolor\fi	
}
\xpatchcmd\@bibitem
{\item}
{\changeBibColor{#1}\item}
{}{\fail}
\xpatchcmd\@lbibitem
{\item}
{\changeBibColor{#2}\item}
{}{\fail}
\makeatother

\begin{document}
	\title{Clutter Suppression, Time-Frequency Synchronization, and Sensing Parameter Association in Asynchronous Perceptive Vehicular Networks}
	
	\author{Xiao-Yang~Wang, Shaoshi~Yang,~\IEEEmembership{Senior~Member,~IEEE}, Jianhua~Zhang,~\IEEEmembership{Senior~Member,~IEEE,}~Christos~Masouros,~\IEEEmembership{Fellow,~IEEE}, Ping~Zhang,~\IEEEmembership{Fellow,~IEEE}
		\thanks{
			This work was supported in part by the Beijing Municipal Natural Science Foundation (No. Z220004), in part by the Beijing Municipal Science \& Technology Commission (No. Z221100007722036), in part by the National Key R\&D Program of China (No. 2023YFB2904803), in part by the Guangdong Major Project of Basic and Applied Basic Research (No.  2023B0303000001), in part by the Fundamental Research Funds for the Central Universities (No. 2023ZCJH02), in part by	the Smart Networks and Services Joint Undertaking (SNS JU) under the European Union’s Horizon Europe Research and Innovation Programme (No. 101139176), and in part by BUPT Excellent Ph.D. Students Foundation (No. CX2023238). \textit{Corresponding author: Shaoshi Yang}.	
			
			X.-Y. Wang and S. Yang are with the School of Information and Communication Engineering, Beijing University of Posts and Telecommunications, and the Key Laboratory of Universal Wireless Communications, Ministry of Education, Beijing 100876, China. Xiao-Yang Wang is also with the Department of Electronic and Electrical Engineering, University College London, London WC1E 7JE, UK (E-mail:  wangxy\_028@bupt.edu.cn, shaoshi.yang@bupt.edu.cn).
			
			C. Masouros is with the Department of Electronic and Electrical Engineering, University College London, London WC1E 7JE, UK (E-mail: c.masouros@ucl.ac.uk).
			
			J. Zhang and P. Zhang are with the School of Information and Communication Engineering, Beijing University of Posts and Telecommunications, and the State Key Laboratory of Networking and Switching Technology, Beijing 100876, China (E-mail: jhzhang@bupt.edu.cn, pzhang@bupt.edu.cn).
		}
	}

	\markboth{Accepted to publish on IEEE Journal on Selected Areas in Communications, vol. 42. no. 10, Oct. 2024.}%
	{Shell \MakeLowercase{\textit{et al.}}: Bare Demo of IEEEtran.cls for IEEE Journals}


	\maketitle
	
	\begin{abstract}
		Significant challenges remain for realizing precise positioning and velocity estimation in practical perceptive vehicular networks (PVN) that rely on the emerging integrated sensing and communication (ISAC) technology. Firstly, complicated wireless propagation environment generates undesired clutter, which  degrades the vehicular sensing performance and increases the computational complexity. Secondly, in practical PVN, multiple types of parameters individually estimated are not well associated with  {specific} vehicles, which may cause error propagation in multiple-vehicle positioning. Thirdly, radio transceivers in a PVN are naturally asynchronous, which causes strong  {range and velocity} ambiguity in vehicular sensing. To overcome these challenges, in this paper  1) we introduce a moving target indication (MTI) based joint clutter suppression and sensing algorithm, and analyze its clutter-suppression performance and the Cramér–Rao lower bound (CRLB) of the paired range-velocity estimation upon using the proposed clutter suppression algorithm;  2) we design an algorithm  {(and its low-complexity versions)} for associating individual direction-of-arrival (DOA) estimates with the paired range-velocity estimates based on ``domain transformation''; 3) we propose the first viable carrier frequency offset (CFO) and time offset (TO) estimation algorithm that supports passive vehicular sensing in non-line-of-sight (NLOS) environments. This algorithm treats the delay-Doppler spectrum of the signals reflected by static objects as an environment-specific ``fingerprint spectrum'', which is shown to exhibit a circular shift property upon changing the CFO and/or TO. Then, the CFO and TO are efficiently estimated by acquiring the number of circular shifts, and we also analyse the mean squared error (MSE) performance of the proposed time-frequency synchronization algorithm.  Finally, simulation results demonstrate the performance advantages of our algorithms under diverse configurations, while corroborating the theoretical analysis.
		
	\end{abstract}
	
	\begin{IEEEkeywords}
		Positioning, clutter suppression, integrated sensing and communication,  {synchronization}, vehicular network. 
	\end{IEEEkeywords}

	%
	\IEEEpeerreviewmaketitle

	\section{Introduction}
	
	\IEEEPARstart{F}{uture} intelligent transportation systems (ITS) are expected to be empowered by advanced wireless networking technologies\cite{Feng_vehicle_offloading}, thus giving birth to the concepts of cellular-vehicle-to-everything (C-V2X) systems and vehicular ad hoc networks (VANETs). 
	
	As cellular networks evolve towards the sixth generation (6G), the demands for precise positioning and velocity estimation of vehicles are rapidly increasing  {\cite{10226306}}. To meet these demands, several solutions can be utilized, including the global navigation satellite systems (GNSS), the distributed cooperative wireless positioning \cite{Lv_coop_localization,cao_Globecom,9868166,Cao_CL_2023} and the cellular network based vehicular sensing\cite{9246715}. In particular, the prospect of exploiting millimetre wave (mmWave) or terahertz (THz) bands for jointly realizing the functions of radar (e.g.,  positioning and velocity estimation) and wireless communications in 6G systems has stimulated substantial interests in the cellular network based solution. By combining the cellular network, the vehicular network and the emerging integrated sensing and communications (ISAC) technology together, we conceive the concept of perceptive vehicular network (PVN) for the future ITS or connected automated vehicles (CAV). In fact, PVN is a particular use case of the general concept of perceptive mobile network (PMN) \cite{zhang2021enabling}, which constitutes a large-scale high-resolution sensing network as the number of base stations (BSs) and the number of remote radio units (RRUs) of each BS are increased \cite{9296833}. As a benefit,  wireless sensing capabilities, e.g., target detection, positioning and tracking, can be embedded in future mobile communication networks  {\cite{9858656,ZJH_ISAC}}.

	In general, there are two types of sensing in a PVN, namely passive sensing and active sensing \cite{9296833}. In passive sensing, the positions and velocities of vehicles are estimated by exploiting the reflected signals that are received at RRUs but originally transmitted by user terminals (UTs) (i.e., uplink passive sensing) or by other RRUs (i.e., downlink passive sensing). By contrast, in active sensing the signals are transmitted by an RRU and received by itself. Nevertheless, high-performance joint communication and active sensing requires full-duplex RRUs. Unfortunately, this has not been mature enough in cellular networks \cite{8999605}. Meanwhile, time-division duplex (TDD) or frequency-division duplex (FDD) based joint communication and sensing \cite{8999605} degrades the achievable communication rate \cite{zhang2021overview}. As a result, in this paper we consider the more practical passive sensing. In addition, since the sensing function needs dedicated data to act as the ``sensing pilot"\footnote{This role can be played by reference signal dedicated for sensing, by the	data payload, or by the existing reference signals designed for other purposes.}, and practical signal transmission protocols between RRUs may be complicated for downlink sensing, we consider the uplink passive sensing scenario as an example herein for convenience of analysis.
	
	Significant challenges remain for realizing passive sensing in a practical PVN. Firstly, the clutter caused by complicated wireless propagation environment in a PVN may substantially increase the computational complexity and degrade the performance of vehicular sensing \cite{9203851,rahman2019framework}. Secondly, in passive vehicular sensing, multiple types of parameters estimated, such as the direction of arrival (DOA),  {the range and the velocity corresponding to each perceived target}, are not well associated with individual vehicles, which may cause error propagation in multiple-vehicle positioning \cite{8936891}. Thirdly, the radio transceivers in a PVN are naturally asynchronous, which causes strong ambiguity in vehicular sensing \cite{ni2021uplink}. 
	
	As far as the first challenge is concerned, the clutter can be regarded as interference\cite{zhang2021enabling}. So far, in the context of PVN, there are two clutter suppression methods relying on exploiting the channel that is associated with the signals reflected by long-term or permanently static objects, namely the recursive moving averaging (RMA) based algorithm \cite{rahman2019framework} and the Gaussian mixture model (GMM) \cite{9203851} based algorithm. 
	The RMA based algorithm eliminates clutter by setting a small forgetting factor to recursively average the received signal. However, it requires a massive number of samples and excessively long time to suppress the clutter.  
	Meanwhile, the GMM based algorithm assumes that the variance of clutter is near zero, while the desired signals have much larger variances. Hence, by learning the mean and variances of different signal components, the statistical GMM model can identify clutter and remove it. Compared with the RMA based algorithm, the GMM based algorithm suppresses the clutter by a smaller number of samples. However, due to the inherent model complexity, it still imposes much higher computational complexity than RMA.

	To overcome the second challenge, in the multiple-vehicle scenario, the estimates of different parameters of the same individual target vehicle must be identified and associated with each other. For example, the individual DOA estimates and the corresponding paired range-velocity estimates \cite{8936891, 7097001, ni2021uplink} must be associated with the appropriate target vehicles. The authors of \cite{8936891} associated the DOA estimates with different UTs by exploiting the UT-specific pseudo-random codes based correlation operation.  
	However, in the context of passive sensing, this method is not able to associate the DOA estimates with the vehicular targets of a PVN. In \cite{7097001}, the DOA was estimated by exploiting both time-domain and spatial-domain measurements, hence the estimates of DOA and the range parameters can be naturally associated with proper UTs\footnote{It is straightforward to convert the values of the range-velocity parameters to the values of the delay-Doppler parameters.}. Nevertheless, when multiple ranges of different UTs are close to each other, some DOAs may be missing. This problem can be solved by invoking the framework of the mirrored multiple signal classification (mirrored-MUSIC) algorithm, if the line-of-sight (LOS) scenario is considered\cite{ni2021uplink}. 
	
	As for the third challenge, effective time-frequency synchronization algorithms are indispensable for estimating and compensating the CFO and TO caused by the oscillators of the spatially isolated transmitters and receivers\cite{Wang_TVT_2023}. However, in the context of PVN, there has been a scarcity of contributions  reported in the open literature on this issue\cite{zhang2021overview}. Among them, the first one is based on cross-antenna cross-correlation (CACC), which was applied to passive Wi-Fi sensing\cite{IndoTrack}. 
	Specifically, the algorithm exploits cross-correlation between signals received at different antennas to extract CFO or TO information. However, the cross-correlation doubles the unknown parameters to be estimated, which results in high computational complexity. Furthermore, for reducing the computational complexity, the authors of \cite{ni2021uplink} proposed a mirrored-MUSIC algorithm, which reconstructs the output of CACC and reduces the number of unknown parameters by half. However, both CACC and mirrored-MUSIC are only applicable in the LOS scenario. 
	Another two contributions, namely FarSense \cite{FarSense} and MultiSense \cite{zeng2020multisense},  applied the channel state information (CSI)-ratio-based CFO and TO synchronization algorithms to the non-line-of-sight (NLOS) scenario.  Unfortunately, they can only synchronize the phase offset between transceivers, which means that it can mitigate the velocity ambiguity, but cannot estimate the delay and thus cannot carry out ranging for target vehicles. Finally, in \cite{9904500} a family of CSI-ratio-based frequency synchronization schemes were proposed, which can only mitigate CFO. Moreover, they are valid only in the scenario where the velocities of the targets are low and the multipath channel between any transceivers is composed of a single dynamic path and multiple static paths.  
	
	\begin{table*}	\newcommand{\tabincell}[2]{\begin{tabular}{@{}#1@{}}#2\end{tabular}}  
		\small
		\centering
		\caption{Comparison between state-of-the-art CFO and/or TO estimation methods and our scheme for asynchronous sensing in PVN}
		\scalebox{1}{\begin{tabular}{c|c|c|c|c}  	
				\hline	
				\textbf{Methods}  & \tabincell{c}{\textbf{Propagation} \\ \textbf{Environment}} & \textbf{Number of Targets} & \tabincell{c}{\textbf{Sensing Functions} \\ \textbf{ Supported}} &\textbf{Number of Antennas}\\ \hline
				\tabincell{c}{CACC \cite{IndoTrack}, \\ mirrored-MUSIC \cite{ni2021uplink}} & LOS & {Multi}-target & velocity, range & Multi-antenna\\ \hline
				FarSense \cite{FarSense} & LOS / NLOS & Single-target&velocity & Multi-antenna\\ \hline
				MultiSense \cite{zeng2020multisense}& LOS / NLOS & Multi-target & velocity & Multi-antenna\\ \hline
				Schemes in \cite{9904500}& LOS / NLOS & Single-target & velocity & Multi-antenna\\ \hline
				Our scheme & LOS / NLOS & Multi-target & velocity, range & \tabincell{c}{Single-antenna,\\ Multi-antenna}\\ \hline
		\end{tabular}} 
		\label{table1}
	\end{table*}

	To overcome the drawbacks of the state-of-the-art solutions conceived for the above-mentioned three challenges, in this paper we study the clutter suppression, time-frequency synchronization, and sensing parameter association problems for DOA, range and velocity estimation in {practical} asynchronous PVNs from a systematic perspective. 
	
	Our novel contributions are summarized as follows, and the major characteristics of the above reviewed schemes and our proposed scheme are summarized in Table~\ref{table1} for clarity.
	\begin{itemize}
		\item[$\bullet$]
		We introduce a single-delay canceller moving target indication (MTI) based clutter suppression algorithm for passive sensing in the orthogonal frequency-division multiplexing (OFDM) aided  {PVN.  Furthermore}, we derive the Cramér–Rao lower bound (CRLB) of the paired range-velocity estimation in the OFDM aided multi-input multi-output (MIMO) passive sensing systems, upon using the proposed clutter suppression scheme in a clutter-rich PVN. Our analysis demonstrates that the CRLB of the MTI based paired range-velocity estimation is related to the particular value of the selected single-delay in MTI, and it is capable of achieving the CRLB of the clutter-free sensing scheme by adjusting the value of the selected single-delay.
	\end{itemize}
	\begin{itemize}
		\item[$\bullet$]
		We develop an algorithm for properly associating the individual DOA estimates with the paired range-velocity estimates. It is valid even if different target vehicles share the same values of the paired range-velocity estimates. Specifically, our algorithm utilizes a spatial filter to pre-process the received signals, and then a domain transformation is invoked to find out the association between the individual DOA estimates and the paired range-velocity estimates. Moreover, two reduced-complexity versions of the algorithm are designed by invoking a simpler low-dimensional domain transformation. The realization of the complexity reduction comes at the expense of  marginal precision degradation of the parameters association.

	\end{itemize}
	\begin{itemize}
		\item[$\bullet$]
		To the best of our knowledge, we propose the first viable joint CFO-TO estimation algorithm for passive sensing in an asynchronous PVN with single-antenna or multi-antenna transceivers operating in NLOS environments.  
		Since there are long-term or permanently static objects in the environment, we treat the delay-Doppler spectrum of the particular signals reflected from the static objects as an environment-specific \textit{fingerprint spectrum}, which is shown to exhibit a circular shift property upon changing the CFO and TO. Thus, by cross-correlation between the fingerprint spectra at different time instants, CFO and TO can be  {estimated.  Finally}, the mean squared error (MSE) of  {the estimates} given by the proposed algorithm is derived. 
	\end{itemize}
	
	The rest of this paper is organized as follows. The system model of an asynchronous PVN relying on the MIMO-OFDM technique is described in Section \uppercase\expandafter {\romannumeral2}.  {Our clutter suppression scheme in the context of passive sensing assuming the ideal synchronous configuration is introduced in Section \uppercase\expandafter {\romannumeral3}.} Then we develop the joint CFO-TO estimation algorithm for asynchronous PVN in Section \uppercase\expandafter {\romannumeral4}. In Section \uppercase\expandafter {\romannumeral5} we derive the CRLB of the paired range-velocity estimation algorithm upon using the proposed clutter suppression scheme, and the MSE performance of the joint CFO-TO estimation algorithm. In Section \uppercase\expandafter {\romannumeral6}, numerical simulations are conducted to consolidate our theoretical analysis and demonstrate the performance of our algorithms. Finally, conclusions are drawn in Section \uppercase\expandafter {\romannumeral7}.  
	
	\textit{Notations}:  We use lower-case and upper-case boldface letters to represent vectors and matrices, respectively. $\lfloor\cdot\rfloor$ denotes the floor function. ${\bf A}^{\textrm T}, {\bf A}^{*}, {\bf A}^{\textrm H}$ and ${\bf A}^{-1}$ represent transpose, conjugate, conjugate transpose and inverse of ${\bf A}$, respectively.  ${\textrm {Re}}({\bf A})$, ${\textrm {Im}}({\bf A})$,  {${\bf A}[i,:]$, ${\bf A}[:,j]$ and ${A}[i,j]$ stands for the real part, the imaginary part, the $i$th row,  the $j$th column, and the $(i,j)$th element of ${\bf A}$, respectively. Moreover, ${\textrm {diag}}({\bf a}_1,\cdots,{\bf a}_n)$ is a block diagonal matrix whose diagonal blocks are $\{{\bf a}_1,\cdots,{\bf a}_n\}$,  {while ${\textrm {diag}}({\bf A})$ represents the vector composed of the diagonal elements of ${\bf A}$}. ${\bf I}_N$ is the $N$-dimensional identity matrix.  {${\textrm {Int}}(\cdot)$ and ${\textrm {Frac}}(\cdot)$ represent the integer and fractional part of a real number, respectively. In addition, ${\mathrm E}(\cdot)$ represents the expectation operator.} Finally, $\oplus$, $\odot$ and $\circledast$ are the right circular shift operator, the Hadamard product operator and the circular convolution operator, respectively.

		\begin{table}[h]
			\scriptsize
			\caption{Definitions of symbols in system model.}
			\label{tab:1}       
			\centering
			\begin{tabularx}{0.48\textwidth} {
					>{\hsize=.35\hsize\linewidth=\hsize}X
					|>{\hsize=1.5\hsize\linewidth=\hsize}X} 
				\hline 
				Symbols&Definitions\\
				\hline
				$L_\textrm{V}$, $L$&The number of targets, and the number of propagation paths.\\
				\hline
				$M_\textrm{R}$, $M_\textrm{U}$&The number of RRU antennas, and the number of UT antennas.\\
				\hline
				$\overset{\rightarrow}{d_1}$, $\overset{\rightarrow}{d_2}$&The line between {the} UT and the $l$th target, and the line between {the} RRU and the $l$th target.\\
				\hline
				$R_{1,l}$, $R_{2,l}$&The  distance {of} the line $\overset{\rightarrow}{d_1}$, and the distance of the line $\overset{\rightarrow}{d_2}$.\\
				\hline
				$v_{l}$&The velocity of the $l$th {target}.\\
				\hline
				$\psi$&{The} angle between the normal direction and {either} the line $\overset{\rightarrow}{d_1}$ or the line $\overset{\rightarrow}{d_2}$.\\
				\hline
				$\tau_\textrm{o}$, $f_\textrm{o}$ &The TO and CFO caused by the separate oscillators of the RRU and the vehicular UT. \\
				\hline
				${\bf H}(t)$&The time-domain channel impulse
				response matrix of the uplink from UT to
				RRU.\\
				\hline
				$h_l$&{The} channel gain of the $l$th path.\\
				\hline
				${\bf a}^{\textrm T}(M_\textrm{R},\phi_{l})$, ${\bf a}^{\textrm T}(M_\textrm{U},\phi_{l})$&The receiving and
				transmitting steering vectors of the $l$th path.\\
				\hline
				$\phi_{l}$, $\theta_{l}$&The DOA and AOD of the $l$th path.\\
				\hline
				$c$&The speed of light.\\
				\hline
				$\tau_{{\textrm d},l}$&$(R_{1,l}+R_{2,l})/c$.\\
				\hline
				$f_\textrm{c}$&The carrier frequency.\\
				\hline
				$\Delta f$&The subcarrier spacing.\\
				\hline
				$T_\textrm{s}$&The sampling
				period in each OFDM symbol.\\
				\hline
				$N_\textrm{s}$&The number of samples in each OFDM symbol.\\
				\hline
				$N_\textrm{cp}$&The number of samples in the CP.\\
				\hline
				$N_\textrm{sub}$&The number of subcarriers.\\
				\hline
				$T_\textrm{sym}$&The OFDM symbol length, $T_\textrm{sym}=N_\textrm{s}T_\textrm{s}$.\\
				\hline
				$p_g(t)$&The $g$th transmitted OFDM symbol of the UT  after excluding the CP.\\
				\hline
				${\bf x}_g$&The data
				symbol vector modulated on the $g$th OFDM
				symbol.\\
				\hline
				${\bf y}_g(t)$&The received {analog bandpass} signal vector.\\
				\hline
				${\bf w}$&The precoding vector.\\
				\hline
				${\bf z}_g(t)$&The zero-mean complex-valued AWGN vector.\\
				\hline
				${\bf z}_g[n]$&The AWGN vector recorded at the $n$th time instant of sampling.\\
				\hline
				$\sigma_0^2$&The power of noise ${\bf z}_g(t)$.\\
				\hline
				$\bar{\bf{y}}_g(t)$&The received analog baseband signal.\\
				\hline
				$\bar{\bf{y}}_g[n]$&The sampled version of $\bar{\bf{y}}_g(t)$ at time instant $nT_\textrm{s}$.\\
				\hline
				$\xi_{\textrm{o}}$, $\xi_{{\textrm D},l}$&The normalized CFO and normalized Doppler frequency shift for the $l$th path.\\
				\hline
				$f_{{\textrm D},l}$&The Doppler frequency shift, $f_{{\textrm D},l} = \frac{2v_l \cos\psi}{c}f_\textrm{c}$.\\
				\hline
				${\bar{\bf y}}_{{\textrm {cp}},g}[n]$&The received baseband signal vector taking into account the extra phase rotation caused by the CP.\\
				\hline
				${{\bf z}}_{{\textrm {cp}},g}[n]$&${{\bf z}}_{{\textrm {cp}},g}[n]={{\bf z}}_{g}[n]e^{-j2\pi g(\xi_{{\textrm D},l}+\xi_{\textrm{o}})\frac{N_\textrm{cp}}{N_\textrm{sub}}}$.\\
				\hline
				${\bf Y}_g$&The matrix of OFDM samples received by all the $M_\textrm{R}$ antennas of the RRU.\\
				\hline
				${\bf F}$&The IDFT matrix.\\
				\hline
				$\boldsymbol{\tau}_{l}$&$\boldsymbol{\tau}_{l}=[1,\cdots, e^{-j2\pi (N_\textrm{sub}-1)\Delta f(\tau_{{\textrm d},l}+\tau_\textrm{o})}]$.\\
				\hline
				${\bf Z}_g$&${\bf Z}_g=[{{\bf z}}_{{\textrm {cp}},g}[1],\cdots, {{\bf z}}_{{\textrm {cp}},g}[N_\textrm{sub}]]$.\\
				\hline
			\end{tabularx}%
		\end{table}
		
		\section{System Model}
		We consider an asynchronous PVN relying on the mmWave band and the MIMO-OFDM technique. As shown in Fig.~\ref{Geometric}, the PVN is composed of a distributed BS, a vehicular UT   participating in the wireless communication with the BS, and $L_\textrm{V}$ vehicles serving as the targets of wireless sensing. For the distributed BS, its baseband unit (BBU) is connected via the fronthaul to an RRU, which is equipped with an $M_\textrm{R}$-element uniform linear array (ULA). The vehicular UT is also equipped with an $M_\textrm{U}$-element ULA. We assume that, in addition to the function of wireless communication, the BS can perform vehicular sensing by utilizing the signals transmitted from the vehicular UT. 
		
		Due to the sparsity of the mmWave channel, we assume that only a small number of propagation paths, denoted by $L$, exist between the vehicular UT and the RRU  \cite{6834753}. The signals propagated via  the $L$ paths can be divided into two categories: 1) the signals reflected by the target vehicles; and 2) the signals reflected by the long-term or permanently static objects in the environment, namely the clutter. Suppose the signal propagated from a particular path, e.g., the $l$th path, experiences reflection by the $l_\textrm{V}$th  target vehicle. Then the distance between the vehicular UT and the $l_\textrm{V}$th target vehicle, namely the length of the line $\overset{\rightarrow}{d_1}$, is $R_{1,l}$, while the distance between the $l_\textrm{V}$th target vehicle and the RRU, namely the length of the line $\overset{\rightarrow}{d_2}$, is $R_{2,l}$. In addition, $v_{l}$ denotes the $l_\textrm{V}$th target vehicle's velocity projected onto the normal direction associated with the $l$th path, and $\psi$ denotes the angle between the normal direction and the line $\overset{\rightarrow}{d_1}$ or the line $\overset{\rightarrow}{d_2}$. We denote the TO and CFO caused by the separate oscillators of the RRU and the vehicular UT as $\tau_\textrm{o}$ and $f_\textrm{o}$, respectively. 
		\begin{figure}[tbp]
			\centering
			\includegraphics[width=3.5in]{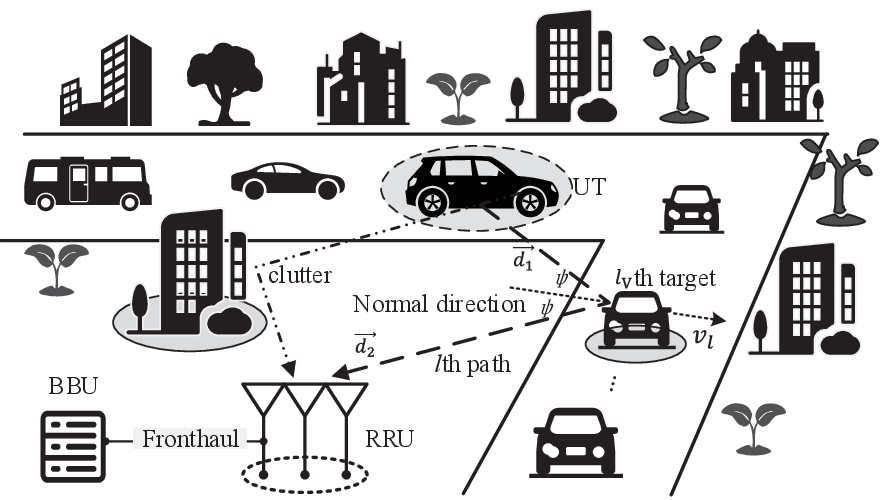}
			\caption{System model of an asynchronous PVN.}\label{Geometric}
		\end{figure}
		
		As a result, the time-domain channel impulse response (CIR) matrix of the uplink from the vehicular UT to the RRU is a function of time $t$ and expressed as\footnote{The CFO is not represented in the CIR function. However, we describe  the CFO in (\ref{equ4_1}).}
		\begin{equation}\label{channel}
			\begin{aligned}
				{\bf H}(t)\! = & {\sum_{l=1}^{L}}h_{l}\delta(t\!-\frac{2v_l \cos\psi}{c}t\!-\!\tau_{{\textrm d},l}\!-\!\tau_\textrm{o}){\bf a}^{\textrm T}(M_\textrm{R},\phi_{l}){\bf a}(M_\textrm{U},\theta_{l}),
			\end{aligned}
		\end{equation}
		where ${\bf a}(M_\textrm{R},\phi_{l}) = [1,\cdots,e^{-\frac{j2\pi (M_\textrm{R}-1)d}{\lambda}\sin\phi_{l}}]$ and ${\bf a}(M_\textrm{U},\theta_{l}) = [1,\cdots,e^{-\frac{j2\pi (M_\textrm{U}-1)d}{\lambda}\sin\theta_{l}}]$ are the receiving and transmitting steering vectors of the $l$th path, respectively. $d$, $\lambda$, $c$ and $\delta(\cdot)$ denote the antenna-element spacing, the signal wavelength, the speed of light and the Dirac delta function, respectively. In addition, $h_{l}$, $\phi_{l}$, $\theta_{l}$, and $\tau_{{\textrm d},l}$ represent the channel gain, the DOA, the angle-of-departure (AOD), and the time-delay of the signal travelling along the $l$th path, respectively, while $\tau_{{\textrm d},l} = (R_{1,l}+R_{2,l})/c$. Moreover, $f_{\textrm c}$ is defined as the carrier frequency.  
		
		{Furthermore, let} us denote the OFDM symbol length as $T_{\textrm {sym}} = N_{\textrm s}T_{\textrm s}$, where $T_{\textrm s}$ and $N_{\textrm s}$ represent the sampling period and the number of samples in each OFDM symbol. Moreover, the number of samples in the cyclic prefix (CP) is denoted by $N_{\textrm {cp}}$, and we define $N_{\textrm {sub}}=N_{\textrm s}-N_{\textrm {cp}}$, which represents the number of subcarriers. Thus, the $g$th transmitted OFDM symbol of the vehicular UT upon excluding the CP is formulated as
		\begin{equation}\label{s_q(t)}
			p_g(t) = e^{j2\pi f_{\textrm c}t}\sum_{u = 0}^{N_\textrm{sub}-1}{x}_g[u]e^{j2\pi u\Delta f t} 
		\end{equation}where we have {${\bf x}_g=[{x}_g[0],\cdots,{x}_g[N_\textrm{sub}-1]]$ and ${x}_g[u]$} is the data symbol modulated on the $u$th subcarrier of the $g$th OFDM symbol, $\Delta f$ is the subcarrier spacing, and $g = 1, \cdots, G$.
		
		In principle, to sense target vehicles, several types of signals, such as the demodulation reference signals (DMRS) \cite{5349184} and the synchronization signal blocks (SSB) defined in 3GPP standards\cite{sadiq2020techniques}, as well as the data payload \cite{rahman2019framework,zhang2021enabling}, can be utilized. Without loss of generality and for achieving high sensing performance with low overhead, in this paper we assume that the data payload is used for sensing the target vehicles.  {Note that our following discussions are also applicable to the scenarios designating DMRS or SSB as the sensing pilots. Then}, the received analog bandpass signal vector ${\bf y}_g(t) = {\bf H}(t)\star {[}{\bf w}p_g(t) {]} + {\bf z}_g(t)$ is further expressed as 
		\begin{equation}\label{y_t}
			\begin{aligned}
				{\bf y}_g(t) &=  {\sum_{l=1}^{L}} h_{l}e^{j2\pi f_\textrm{c}\bar t}{\bf a}^{\textrm T}(M_\textrm{R},\phi_{l})
				{\bf a}(M_\textrm{U},\theta_{l}){\bf w}\\
				&\ \cdot\sum_{u=0}^{N_\textrm{sub}-1}{x}_g[u]e^{j2\pi u\Delta f\bar t} +{\bf z}_g(t),
			\end{aligned}
		\end{equation}
		where ``$\star$'' denotes convolution, ${\bf w}$ is an ${M_\textrm{U}}$-dimensional column vector of precoding, $\bar t \triangleq t-\frac{2v_l \cos\psi}{c}t-\tau_{{\textrm d},l}-\tau_\textrm{o}$, and ${\bf z}_g(t) {\in\mathbb{C}^{M_{\textrm R}\times 1}}$ is the zero-mean complex-valued additive white Gaussian noise (AWGN) vector satisfying $\textrm{E}[{{\bf z}}_g^*(t){{\bf z}}_g(t)] = \sigma_0^2{{\bf I}_{M_{\textrm{R}}}}$.  
		
		Next, ${\bf{y}}_g(t) {\in\mathbb{C}^{M_{\textrm R}\times 1}}$ is demodulated to a baseband signal by multiplying it with the local carrier, which typically exhibits a frequency offset $f_\textrm{o}$ relative to the transmitter as described in \cite{6760595}, thus yielding 
		\begin{equation}\label{equ4_1}
			\bar{\bf{y}}_g(t) = {\bf{y}}_g(t)e^{-j2\pi(f_\textrm{c}+f_\textrm{o})t}.
		\end{equation} Then, $\bar{\bf{y}}_g(t)$ is sampled at time instant $nT_{\textrm s}$. As a result, the analog signal vector $\bar{\bf{y}}_g(t)$ is transformed into a digital signal vector ${\bar{\bf y}}_g[nT_{\textrm s}], n=1,\cdots,N_{\textrm {sub}}$. For brevity, in what follows ${\bar{\bf y}}_g[nT_{\textrm s}]$ is rewritten as ${\bar{\bf y}}_g[n]$. Note that $\Delta f\frac{2 v_{l}\cos\psi}{c}t$ is small enough on the time scale of the duration of a few hundred OFDM symbols needed for sensing\cite{liu2020super}, while $(f_\textrm{o}+\frac{2 v_{l}f_\textrm{c}\cos\psi}{c})t$ in $\bar{\bf{y}}_g(t)$ is small enough on the time scale of the duration of a single OFDM symbol. Hence, in the duration of  {a few hundred OFDM symbols and} a single OFDM symbol we can approximate $e^{-j2\pi u\Delta f\frac{2v_{l}\cos\psi}{c}t}$ as 1 and $e^{-j2\pi(f_\textrm{o}+\frac{2 v_{l}f_\textrm{c}\cos\psi}{c})t}$ as a constant, respectively\footnote{After sampling, the term $e^{-j2\pi(f_\textrm{o}+\frac{2 v_{l}f_\textrm{c}\cos\psi}{c})t}$ corresponds to the $n$th sample of the $g$th OFDM symbol is $e^{-j2\pi(f_\textrm{o}+\frac{2 v_{l}f_\textrm{c}\cos\psi}{c})((g-1)N_\textrm{sub}T_\textrm{s}+nT_\textrm{s})}$, which is approximated as the constant $e^{-j2\pi(f_\textrm{o}+\frac{2 v_{l}f_\textrm{c}\cos\psi}{c})(g-1)N_\textrm{sub}T_\textrm{s}}$ within the duration of the $g$th OFDM symbol. However, this constant becomes larger when increasing the index $g$ of the OFDM symbol.}. However, when $f_\textrm{o}$ is very large, the traditional coarse CFO synchronization algorithms for OFDM, such as the Moose frequency synchronization algorithm in \cite{Moose328961}, should be implemented to eliminate the term $e^{-j2\pi(f_\textrm{o}+\frac{2 v_{l}f_\textrm{c}\cos\psi}{c})t}$. As a result, ${\bar{\bf y}}_g[n] {\in\mathbb{C}^{M_{\textrm R}\times 1}}$ is approximately expressed as \eqref{equ4},
		\begin{figure*}
			\begin{equation} \label{equ4}
				{\bar{\bf y}}_g[n]
				\!\approx \!  {\sum_{l=1}^{L}}\!\sum_{u=0}^{N_\textrm{sub}-1}\! h_{l} e^{-j2\pi f_{\textrm c} (\tau_{{\textrm d},l}\!+\tau_\textrm{o})} e^{-j2\pi(f_{\textrm o}+\frac{2 v_{l}f_\textrm{c}\cos\psi}{c})(g-1)N_\textrm{sub}T_{\textrm s}}
				{\bf a}^{\textrm T}(M_\textrm{R},\phi_{l}) {\bf a}(M_\textrm{U},\theta_{l}){\bf w} {x}_g[u]e^{-j2\pi u\Delta f(\tau_{{\textrm d},l}\!+\tau_\textrm{o})}e^{j2\pi u\Delta fnT_{\textrm s}}\! + \!{{\bf z}}_g[n],
			\end{equation}
			\hrulefill
		\end{figure*}
		where ${{\bf z}}_g[n]$ is the complex-valued AWGN vector recorded at the $n$th time instant of sampling for the $g$th transmitted OFDM symbol. Additionally, by defining $\xi_{\textrm{o}}=N_\textrm{sub}f_{\textrm{o}}T_{\textrm s}$, $f_{{\textrm D},l} = \frac{2v_l \cos\psi}{c}f_\textrm{c}$, and $\xi_{{\textrm D},l} = N_\textrm{sub}
		f_{{\textrm D},l}T_{\textrm s}$, we can represent the term $e^{-j2\pi(f_{\textrm o}+\frac{2 v_{l}f_\textrm{c}\cos\psi}{c})(g-1) {N_\textrm{sub}T_\textrm{s}}}$ as $e^{-j2\pi (\xi_{{\textrm D},l}+\xi_{\textrm{o}})(g-1)}$. 
		
		It is worth noting that the CP is not characterized in (\ref{s_q(t)}), which is the model of the transmitted OFDM symbol $p_g(t)$. However, in realistic systems the time occupied by transmitting the CP actually causes extra phase rotation on the non-CP part of OFDM symbols. Therefore, when taking into account the extra phase rotation caused by the CP, the received baseband signal vector is reformulated as  ${\bar{\bf y}}_{{\textrm {cp}},g}[n]={\bar{\bf y}}_g[n]e^{-j2\pi g(\xi_{{\textrm D},l}+\xi_{\textrm{o}})\frac{N_{\textrm {cp}}}{N_\textrm{sub}}}$. More specifically, we have
		\begin{equation} \label{newequ4}
			\begin{aligned}
				&{\bar{\bf y}}_{{\textrm {cp}},g}[n]
				\approx   {\sum_{l=1}^{L}}\sum_{u=0}^{N_\textrm{sub}-1} h_{l} e^{-j2\pi f_{\textrm c} (\tau_{{\textrm d},l}+\tau_\textrm{o})} e^{-j2\pi (\xi_{{\textrm D},l}+\xi_{\textrm{o}})\frac{N_\textrm{cp}}{N_\textrm{sub}}}\\
				& \cdot  e^{-j2\pi (\xi_{{\textrm D},l}+\xi_{\textrm{o}})(g-1)\frac{N_{\textrm s}}{N_\textrm{sub}}}{\bf a}^{\textrm T}(M_\textrm{R},\phi_{l})  {\bf a}(M_\textrm{U},\theta_{l}){\bf w} {x}_g[u]\\
				& \cdot e^{-j2\pi u\Delta f(\tau_{{\textrm d},l}+\tau_{\textrm o})}e^{j2\pi u\Delta fnT_{\textrm s}} + {{\bf z}}_{{\textrm {cp}},g}[n],
			\end{aligned}
		\end{equation}
		where ${{\bf z}}_{{\textrm {cp}},g}[n]={{\bf z}}_{g}[n]e^{-j2\pi g(\xi_{{\textrm D},l}+\xi_{\textrm{o}})\frac{N_\textrm{cp}}{N_\textrm{sub}}}$.
		Furthermore, we construct ${\bf Y}_g$ by $[{\bar{\bf y}}_{{\textrm {cp}},g}[1],\cdots, {\bar{\bf y}}_{{\textrm {cp}},g}[N_\textrm{sub}]]$. Here ${\bf Y}_g \in \mathbb{C}^{M_\textrm{R} \times N_\textrm{sub}}$ represents the OFDM samples received by all the $M_\textrm{R}$ antennas of the RRU in the duration of a single OFDM symbol. ${\bf Y}_g$ can be further formulated as 
		\begin{equation}\label{Y_g}
			\begin{aligned}
				&{\bf Y}_g\! \approx\!\!  {\sum_{l=1}^{L}}\! h_{l}  e^{-j2\pi\! (\xi_{{\textrm D},l}\!+\!\xi_{\textrm{o}})\frac{N_\textrm{cp}}{N_\textrm{sub}}} e^{-j2\pi (\xi_{{\textrm D},l}\!+\!\xi_{{\textrm o}})(g-1)\frac{N_{\textrm s}}{N_\textrm{sub}}}
				e^{-j2\pi f_\textrm{c} (\tau_{{\textrm d},l}\!+\!\tau_\textrm{o})}\\
				&\cdot{\bf a}^{\textrm T}(M_\textrm{R},\phi_{l}){\bf a}(M_\textrm{U},\theta_{l}){\bf w} \boldsymbol{\tau}_{l}{\textrm D}({\bf x}_g){\bf F}\!+\! {\bf Z}_g,
			\end{aligned}
		\end{equation}
		where ${\bf F}$, ${\textrm D}({\bf x}_g)$, $\boldsymbol{\tau}_{l}$ and ${\bf Z}_g$ represent the inverse discrete Fourier transform (IDFT) matrix, $\textrm{diag}({\bf x}_g)$, $[1,\cdots, e^{-j2\pi (N_\textrm{sub}-1)\Delta f(\tau_{{\textrm d},l}+\tau_\textrm{o})}]$ and $[{{\bf z}}_{{\textrm {cp}},g}[1],\cdots, {{\bf z}}_{{\textrm {cp}},g}[N_\textrm{sub}]]$,  respectively. 
		
		Since a large number of mathematical symbols are defined when constructing the system model, Table \ref{tab:1} is provided to summarize symbol definitions for ease of understanding the system model.

		\section{Clutter Suppression and Uplink Passive Sensing in Synchronous PVN}
		\begin{figure}[tbp]
			\centering
			\includegraphics[width=3.5in]{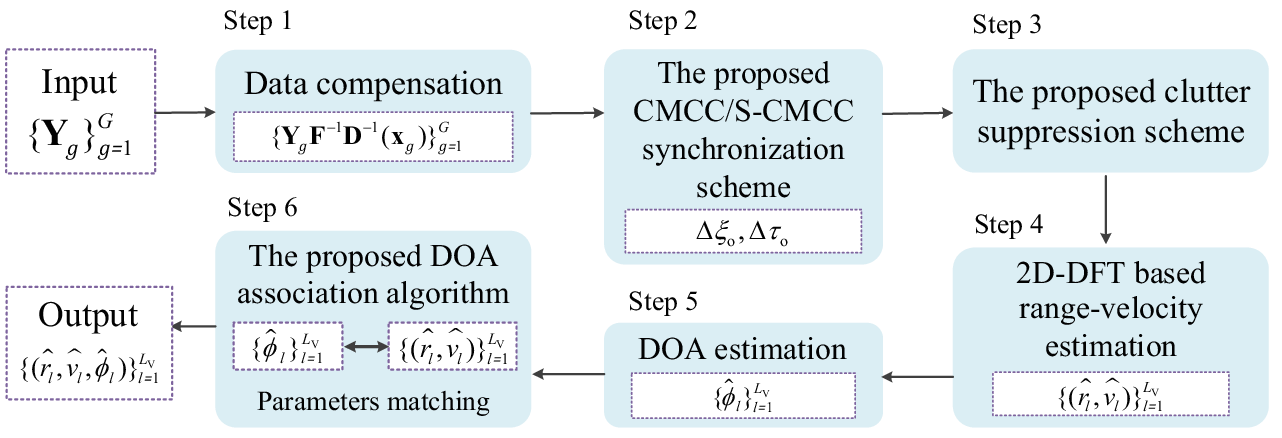}
			\caption{ {Schematic diagram of the whole sensing scheme designed for asynchronous PVN. With the input $\{{\bf Y}_g\}_{g=1}^{G}$,  our scheme is composed of the modules of data compensation, synchronization, clutter suppression, range-velocity estimation, DOA estimation and DOA association. The outputs of the whole scheme are  $L_{\textrm V}$ sets of velocity-range-DOA parameters. The parameters in each of these sets correspond to an individual target vehicle collectively. Moreover, the DOA estimation module at Step 5 can be implemented by invoking any DOA estimation algorithm, such as MUSIC.}}\label{fig2}
		\end{figure}
		For the sake of clarity, the overall framework of our sensing scheme  {designed for practical PVN} is shown in Fig.~\ref{fig2}. {Specifically, the data compensation module in Step 1 has been widely used in  existing studies \cite{zhang2021enabling}. The time-frequency synchronization module in Step 2 will be detailedly introduced in Section \uppercase\expandafter{\romannumeral4}.  Moreover, the clutter suppression and range-velocity estimation modules are constituted by Step 3 and Step 4, respectively, while the DOA estimation and the DOA association are implemented in Step 5 and Step 6, respectively.}
		
		To gain basic but valuable insights into the design of vehicular sensing schemes suitable for {realistic} environment, let us start with assuming an ideally synchronous PVN. In this case, the time-frequency synchronization module in Step~2 is omitted. Then, our scheme can be divided into two main parts: 1) clutter suppression and range-velocity estimation, and 2) DOA estimation and the association of DOA results with the corresponding  {paired} range-velocity parameters.

		\subsection{Joint Clutter Suppression and Range-Velocity Estimation  {Scheme}}\label{sec:clutter_suppression}
		Before {suppressing} the clutter and performing the paired range-velocity estimation, we firstly compensate $\mathbf{Y}_g$ by exploiting methods presented in \cite{liu2020super}.   {Note that the perception module in a PVN exhibits insensitivity to sensing latency on the millisecond timescale, primarily because of the minimal changes in velocity and position of the individual targets in such a short period. Thus,} the demodulated communication data  {${\bf x}_g$} in the communication module, can be shared as reference (or pilot) data with {the perception} module, to compensate $\mathbf{Y}_g {\in\mathbb{C}^{M_{\textrm R}\times N_\textrm{sub}}}$ by 
		\begin{equation}\label{data_compensation}
			\hat{{\bf {Y}}}_g={\bf Y}_g{\bf F}^{-1}{\textrm  D}^{-1}({\bf x}_g),
		\end{equation}which is further formulated as
		\begin{equation} \label{newequ12}
			\begin{aligned}
				\hat{{\bf {Y}}}_g=  {\sum_{l=1}^{L}}\ \bar{h}_{l}\Phi_{l,g}{\bf a}^{\textrm T}(M_\textrm{R},\phi_{l}){\bf a}(M_\textrm{U},\theta_{l}){\bf w} \boldsymbol{\tau}_{l} + \hat{\bf Z}_g,
			\end{aligned}
		\end{equation}where $\bar{h}_{l}=h_{l} e^{-j2\pi f_{\textrm c} \tau_{l}}$, $\Phi_{l,g}=e^{-j2\pi \xi_{l}\frac{N_\textrm{cp}}{N_\textrm{sub}}} e^{-j2\pi \xi_{l}(g-1)\frac{N_{\textrm s}}{N_\textrm{sub}}}$, and $\hat{\bf Z}_g={{\bf Z}}_g{\bf F}^{-1}{\textrm  D}^{-1}({\bf x}_g)$. Note that $\hat{{\bf {Y}}}_g {\in\mathbb{C}^{M_{\textrm R}\times N_\textrm{sub}}}$  essentially reflects the overall characteristics of the \textit{equivalent channel}.  {Here, we define $\xi_{l}=\xi_{{\textrm D},l}+\xi_{\textrm{o}}$ and $\tau_{l}=\tau_{{\textrm d},l}+\tau_{\textrm{o}}$ for convenience. In synchronous systems where neither CFO nor TO exists, we have $\xi_{l}=\xi_{{\textrm D},l}$ and $\tau_{l}=\tau_{{\textrm d},l}$.}
		
		To suppress the clutter, the static characteristics of the CIR associated with the clutter are exploited. According to (\ref{channel}), this CIR is formulated as  
		\begin{equation}\label{clutterchannel}
			\begin{aligned}
				{\bf H}_\textrm{clutter}(t) = & {\sum_{l=1}^{L_1}}h_{l}\delta(t-\tau_{{\textrm d},l}){\bf a}^{\textrm T}(M_\textrm{R},\phi_{l}){\bf a}(M_\textrm{U},\theta_{l}),
			\end{aligned}
		\end{equation}
		where $L_1$ is defined as the number of clutter paths. Obviously, the clutter channel exhibits quasi-constant gains and phases on all the propagation paths. In contrast, the phases of the CIR of the mobile  {signal} components, namely the signals {reflected} by moving objects, change constantly\cite{FarSense,zeng2020multisense}. 
		Thus, {by exploiting different channel characteristics}, the clutter components can be eliminated with the {well known} single-delay-canceller MTI  {scheme\cite{MTI_single-delay-canceller}, yielding  
			\begin{equation} \label{equ18}
				\begin{aligned}
					{\breve{\bf Y}}_g\!=\!& {\hat{\bf Y}}_g - {\hat{\bf Y}}_{g-G_\textrm{d}} \\
					=\!& {\sum_{l=1}^{L}}e^{-j2\pi \xi_{l}(g\!-\!1)\frac{N_{\textrm s}}{N_\textrm{sub}}}(1\!-\!e^{j2\pi \xi_{l}G_\textrm{d}\frac{N_{\textrm s}}{N_\textrm{sub}}})e^{-j2\pi \xi_{l}\frac{N_\textrm{cp}}{N_\textrm{sub}}}\boldsymbol{\alpha}_l\boldsymbol{\tau}_{l} \!+\!\breve{\bf Z}_g,
				\end{aligned}
			\end{equation}
			where $G_\textrm{d}$ is the selected delay in MTI, $\breve{\bf Z}_g = \hat{\bf Z}_g-\hat{\bf Z}_{g-G_\textrm{d}}$, and $\boldsymbol{\alpha}_l = \bar{h}_{l}{\bf a}^{\textrm T}(M_\textrm{R},\phi_{l}) {\bf a}(M_\textrm{U},\theta_{l}){\bf w}$.
			
			{Then, let us define $\breve{\mathbf{y}}_{g,m}$ as the $m$th row of $\breve{\mathbf{Y}}_g$, to represent the $g$th compensated OFDM symbol received by the $m$th antenna of the RRU. $\breve{\mathbf{y}}_{g,m}$ is stacked as ${\boldsymbol\Gamma}_m=[\breve{\mathbf{y}}_{1,m}^{\textrm T},\cdots,\breve{\mathbf{y}}_{G,m}^{\textrm T}]^{\textrm T}$. Without loss of generality, we assume that the signals received by the $m$th antenna have the highest signal-to-noise ratio (SNR) among all the signals received by the $M_\textrm{R}$ antennas of the RRU. This assumption guarantees that the subsequent range-velocity estimation utilizing ${\boldsymbol\Gamma}_m$ performs as well as possible.}
			Specifically, ${\boldsymbol\Gamma}_m$ can be formulated as
			\begin{equation}\label{Gamma}
				{\boldsymbol\Gamma}_m=\sum_{l=1}^{L}\!\alpha_l[m]\boldsymbol{\xi}_l\boldsymbol{\tau}_{l} +\tilde{\bf Z},
			\end{equation}
			where  {$\alpha_l[m]$ is the $m$th element of $\boldsymbol{\alpha}_l$ and} $\boldsymbol{\xi}_l$ is defined as $[(1\!-\!e^{j2\pi \xi_{l}G_\textrm{d}\frac{N_{\textrm s}}{N_\textrm{sub}}})e^{-j2\pi \xi_{l}\frac{N_\textrm{cp}}{N_\textrm{sub}}},\cdots$ $,e^{-j2\pi \xi_{l}(G-1)\frac{N_{\textrm s}}{N_\textrm{sub}}}$ $(1\!-\!e^{j2\pi \xi_{l}G_\textrm{d}\frac{N_{\textrm s}}{N_\textrm{sub}}})e^{-j2\pi \xi_{l}\frac{N_\textrm{cp}}{N_\textrm{sub}}}] {^{\textrm T}}$.  {Moreover,} we have $\tilde{\bf  Z}=[\breve{\mathbf{z}}_{1,m}^{\textrm T},\cdots,\breve{\mathbf{z}}_{G,m}^{\textrm T}]^{\textrm T}$, {where $\breve{\mathbf{z}}_{g,m}$ is the $m$th row of $\breve{\bf Z}_g$}.
			
			{According to (\ref{Gamma}), it is intuitive that $\boldsymbol\Gamma_m$ can be deemed as the linear weighted sum of $L$ two-dimensional (2D) discrete single-tone signal sequence ${\boldsymbol\xi}_{l}{\boldsymbol\tau}_{l}$, $l=1,\cdots,L$, subject to an AWGN term.}
			As a result, the 2D-DFT can be applied to perform frequency analysis of $\boldsymbol{\Gamma}_m$. It is obvious that one can estimate the desired range-velocity parameters by locating the peaks of the frequency spectrum. However, performing 2D-DFT can result in high computational complexity. To reduce the complexity, we intend to perform 2D-DFT on the real part of $\boldsymbol{\Gamma}_m$, namely ${\textrm {Re}}(\boldsymbol{\Gamma}_m)$, and obtain ${\bf \bar Y}={\bf F}_G^{\textrm H}{\textrm {Re}}(\boldsymbol{\Gamma}_m){\bf F}_{N_\textrm{sub}}^{\textrm H}$.  Since ${\textrm {Re}}(\boldsymbol{\Gamma}_m)$ has the same SNR as $\boldsymbol{\Gamma}_m$, this operation does not affect the estimation accuracy. 
			
			Furthermore, in practice the frequency spectrum peaks of ${\textrm {Re}}(\boldsymbol{\Gamma}_m)$ may not exactly locate on the grid of 2D-DFT, which is sampled from the discrete time Fourier transform (DTFT) of ${\textrm {Re}}(\boldsymbol{\Gamma}_m)$ \cite{oppenheim2001discrete}. As a result, to find the deterministic relationship between the angular frequencies of ${\boldsymbol\xi}_{l}{\boldsymbol\tau}_{l}$ and its corresponding peak coordinates within the $l$th component matrix\footnote{The composition of ${\bf \bar Y}$ is given by \eqref{delayspectrum}.} of ${\bf \bar Y}$, 
			$\forall l=1,\cdots,L$,  it is necessary to further derive the expression of ${\bf \bar Y}$.
			
			Firstly, according to (\ref{Gamma}), the $( {k},n)$th element of the noise-free ${\textrm {Re}}(\boldsymbol{\Gamma}_m)$ can be represented as 
			{\begin{equation}\label{Re(gamma)}
					\begin{aligned}
						&\sum_{l=1}^{L}\left\{{\cos}(2\pi n\Delta f\tau_{l}) {\textrm {Re}}\{\alpha_l[m]e^{-j2\pi \xi_{l}(k-1)\frac{N_{\textrm s}}{N_\textrm{sub}}}(1-e^{j2\pi \xi_{l}G_\textrm{d}\frac{N_{\textrm s}}{N_\textrm{sub}}})\right.\\
						&\left.e^{-j2\pi \xi_{l}\frac{N_\textrm{cp}}{N_\textrm{sub}}}\}\!+\!{\sin}(2\pi n\Delta f\tau_{l}){\textrm {Im}}\{\alpha_l[m]e^{-j2\pi \xi_{l}(k\!-\!1)\frac{N_{\textrm s}}{N_\textrm{sub}}}\right.\\
						&\left.(1-e^{j2\pi \xi_{l}G_\textrm{d}\frac{N_{\textrm s}}{N_\textrm{sub}}})e^{-j2\pi \xi_{l}\frac{N_\textrm{cp}}{N_\textrm{sub}}}\}\right\}.
					\end{aligned}
			\end{equation}}
			
			Moreover, according to (\ref{Re(gamma)}), ${\textrm {Re}}(\boldsymbol{\Gamma}_m)$ can be concisely expressed as  
			\begin{equation}\label{realgamma}
				{\textrm {Re}}(\boldsymbol{\Gamma}_m) = \sum_{l=1}^{L}\boldsymbol{\Lambda}_l\boldsymbol{\Theta}_l+{\textrm {Re}}(\tilde{\bf  Z}),
			\end{equation}
			where the $n$th column of $\boldsymbol{\Theta}_l$ is $[{\cos}(2\pi n\Delta f\tau_{l}), {\sin}(2\pi n\Delta f\tau_{l})]^\textrm{T}$
			and the $ {k}$th row of $\boldsymbol{\Lambda}_l$ is
			\begin{equation}\label{lambda}
				\begin{aligned}
					&\left[{\textrm {Re}}\{\alpha_l[m]e^{-j2\pi \xi_{l}(k-1)\frac{N_{\textrm s}}{N_\textrm{sub}}}(1\!-\!e^{j2\pi \xi_{l}G_\textrm{d}\frac{N_{\textrm s}}{N_\textrm{sub}}})e^{-j2\pi \xi_{l}\frac{N_\textrm{cp}}{N_\textrm{sub}}}\},\ \right.\\
					&\left.{\textrm {Im}}\{\alpha_l[m]e^{-j2\pi \xi_{l}(k-1)\frac{N_{\textrm s}}{N_\textrm{sub}}}(1\!-\!e^{j2\pi \xi_{l}G_\textrm{d}\frac{N_{\textrm s}}{N_\textrm{sub}}})e^{-j2\pi \xi_{l}\frac{N_\textrm{cp}}{N_\textrm{sub}}}\} \right].
				\end{aligned}
			\end{equation}
			
			Then, the 2D-DFT  {of ${\textrm {Re}}(\boldsymbol{\Gamma}_m)$}  can be represented as
			\begin{equation}\label{2D-DFT-1}
				{{\bf \bar Y}} = \sum_{l=1}^{L}{\bf F}_G^{\textrm H}\boldsymbol{\Lambda}_l({\bf F}_{N_\textrm{sub}}^{\textrm H}\boldsymbol{\Theta}^{\textrm T}_l)^{\textrm T}+\bar{{\bf Z}},
			\end{equation}where $\bar{{{\bf Z}}} = {\bf F}_G^{\textrm H}{\textrm {Re}}(\tilde{\bf Z}){\bf F}_{N_\textrm{sub}}^{\textrm H}$. 
			Furthermore, by leveraging the convolution theorem, ${\bf F}_G^{\textrm H}\boldsymbol{\Lambda}_l[:,n],\ n=1,2$, is expressed as 
			\begin{equation}\label{equ15}
				{\bf F}_G^{\textrm H}\boldsymbol{\Lambda}_l[:,n]={\bf F}_G^{\textrm H}{\bf I}_{G}\boldsymbol{\Lambda}_l[:,n]=({\bf F}_G^{\textrm H}{\textrm {diag}}({\bf I}_{G}))\circledast \boldsymbol{\gamma}_{G,n},
			\end{equation}
			where the vector, ${\bf F}_G^{\textrm H}{\textrm {diag}}({\bf I}_{G})\in\mathbb{C}^{G\times 1}$, is the DFT of the discrete rectangular window function of length-$G$ in samples, and $\boldsymbol{\gamma}_{G,n}\in\mathbb{C}^{G\times 1},\ n=1,2$, represents the DFT of the periodically extended sequence of $\boldsymbol{\Lambda}_l[:,n]$ \cite{oppenheim2001discrete}. Specifically, ${\bf F}_G^{\textrm H}{\textrm {diag}}({\bf I}_{G})\in\mathbb{C}^{G\times 1}$ can also be obtained by sampling the output of the DTFT of the length-$G$ discrete rectangular window function, and this DTFT is expressed as
			\begin{equation}
				{S}_{G}(f)=\sum_{g=0}^{G-1}e^{-j2\pi fg}.
			\end{equation} 
			
			It is well known that ${S}_{G}(\cdot)$ is an aliased sinc function\cite{oppenheim2001discrete}.
			Moreover, according to the expression of $\boldsymbol{\Lambda}_l$, we can formulate the $k$th row of $[\boldsymbol{\gamma}_{G,1},\boldsymbol{\gamma}_{G,2}]$ as
			\begin{equation}\label{DFT3}
				\begin{aligned}
					\frac{1}{2}
					&\left( {\textrm {Re}}\{  \tilde{\alpha}_l[m]\}\{\delta(k+\xi_{l}\frac{N_{\textrm s}}{N_\textrm{sub}})+\delta(k-\xi_{l}\frac{N_{\textrm s}}{N_\textrm{sub}})\} \right. \\
					& -j{\textrm {Im}}\{\tilde{\alpha}_l[m]\}\{\delta(k+\xi_{l}\frac{N_{\textrm s}}{N_\textrm{sub}})-\delta(k-\xi_{l}\frac{N_{\textrm s}}{N_\textrm{sub}})\}, \\
					&j{\textrm {Re}}\{\tilde{\alpha}_l[m]\}\{\delta(k\!+\!\xi_{l}\frac{N_{\textrm s}}{N_\textrm{sub}})\!-\!\delta(k-\xi_{l}\frac{N_{\textrm s}}{N_\textrm{sub}})\} \\
					&\left. +{\textrm {Im}}\{\tilde{\alpha}_l[m]\}\{\delta(k+ \xi_{l}\frac{N_{\textrm s}}{N_\textrm{sub}})+\!\delta(k \!-\!\xi_{l}\frac{N_{\textrm s}}{N_\textrm{sub}})\} \right),
				\end{aligned}
			\end{equation}
			where $\tilde{\alpha}_l[m]={\alpha}_l[m](1\!-\!e^{j2\pi \xi_{{\textrm D},l}G_\textrm{d}\frac{N_{\textrm s}}{N_\textrm{sub}}})e^{-j2\pi \xi_{{\textrm D},l}\frac{N_\textrm{cp}}{N_\textrm{sub}}}$.
			According to (\ref{equ15}), the expression of the ${k}$th row of ${\bf F}_G^{\textrm H}\boldsymbol{\Lambda}_l$ is   
			\begin{equation}\label{DFT1}
				\begin{aligned}
					\frac{1}{2}
					&\left({\textrm {Re}}\{ \tilde{\alpha}_l[m]\}\{{S}_{G}[(k+\xi_{l}\frac{N_{\textrm s}}{N_\textrm{sub}})f_{\textrm R}]+{S}_{G}[(k-\xi_{l}\frac{N_{\textrm s}}{N_\textrm{sub}})f_{\textrm R}]\} \right.\\
					&-j{\textrm {Im}}\{\tilde{\alpha}_l[m]\}\{{S}_{G}[(k+\xi_{l}\frac{N_{\textrm s}}{N_\textrm{sub}})f_{\textrm R}]-{S}_{G}[(k-\xi_{l}\frac{N_{\textrm s}}{N_\textrm{sub}})f_{\textrm R}]\},\\
					&j{\textrm {Re}}\{\tilde{\alpha}_l[m]\}\{{S}_{G}[(k\!+\!\xi_{l}\frac{N_{\textrm s}}{N_\textrm{sub}})f_{\textrm R}]\!-\!{S}_{G}[(k-\xi_{l}\frac{N_{\textrm s}}{N_\textrm{sub}})f_{\textrm R}]\}\\
					&\left.+{\textrm {Im}}\{\tilde{\alpha}_l[m]\}\{{S}_{G}[(k+ \xi_{l}\frac{N_{\textrm s}}{N_\textrm{sub}})f_{\textrm R}]+\!{S}_{G}[(k \!-\!\xi_{l}\frac{N_{\textrm s}}{N_\textrm{sub}})f_{\textrm R}]\} \right),
				\end{aligned}
			\end{equation}
			where $f_{\textrm R}$ is the frequency resolution when DFT is obtained by sampling DTFT. In particular,  $f_{\textrm R}$ can be expressed as the reciprocal of the total sensing time\cite{liu2020super}, i.e., $\frac{1}{GT_{\textrm {sym}}}$.
			
			Similarly, the $n$th column of $\boldsymbol{\Theta}_l{\bf F}_{N_\textrm{sub}}^*$ can be formulated as 
			\begin{equation}\label{DFT2}
				\begin{aligned}
					&\frac{1}{2}\left( {S}_{N_\textrm{sub}}[( n+\frac{ \tau_l}{T_{\textrm s}})T_{\textrm R}]+{S}_{N_\textrm{sub}}[( n-\frac{ \tau_{l}}{T_{\textrm s}})T_{\textrm R}],\right.\\
					&\left. -j\{ {S}_{N_\textrm{sub}}[( n+\frac{ \tau_l}{T_{\textrm s}})T_{\textrm R}]-{S}_{N_\textrm{sub}}[( n-\frac{ \tau_l}{T_{\textrm s}})T_{\textrm R}]\}\right)^{\textrm T},
				\end{aligned}
			\end{equation} where  {$T_{\textrm R}$ is the time resolution quantified as the reciprocal of the total sensing bandwidth\cite{liu2020super}, i.e., $\frac{1}{N_\textrm{sub}\Delta f}$. Moreover,} ${S}_{N_\textrm{sub}}$  is defined as the DTFT of the length-${N_\textrm{sub}}$ discrete rectangular window function\cite{oppenheim2001discrete}.  From (\ref{DFT1}) and (\ref{DFT2}), it is obvious that the individual columns of ${\bf F}_G^{\textrm H}\boldsymbol{\Lambda}_l$ and individual rows of $\boldsymbol{\Theta}_l{\bf F}_{N_\textrm{sub}}^*$ are both based on {discrete samples of aliased sinc functions}. 
			
			Then, by substituting (\ref{DFT1}) and (\ref{DFT2}) into (\ref{2D-DFT-1}), we obtain the final expression of the $(k,n)$th element of ${\bf \bar Y}$ as
			\begin{equation}\label{delayspectrum}
				\begin{aligned}
					&{\bar Y}[k,n]=\frac{1}{4}\sum_{l=1}^{L}\left\{\tilde{\alpha}_l[m]{S}_{G}[(k-\xi_{l}\frac{N_{\textrm s}}{N_\textrm{sub}})f_{\textrm R}] 
					{S}_{N_\textrm{sub}}[( n-\frac{ \tau_l}{T_{\textrm s}})T_{\textrm R}]\right.\\
					&+\left.\tilde{\alpha}_l^*[m]{S}_{G}[(k+\xi_{l}\frac{N_{\textrm s}}{N_\textrm{sub}})f_{\textrm R}] 
					{S}_{N_\textrm{sub}}[( n+\frac{ \tau_l}{T_{\textrm s}})T_{\textrm R}]\right\}+\bar{{Z}}[k,n],\\
				\end{aligned}
			\end{equation}where $\bar{{Z}}[k,n]$ is the $(k,n)$th element of $\bar{{\bf Z}}$.
			According to (\ref{delayspectrum}), each element of ${\bf \bar Y}$ is the linear weighted sum of multiple discrete 2D aliased sinc functions.  Firstly, \textcolor{black}{as observed from (\ref{delayspectrum}), corresponding to a given $l$, the absolute value of the component output matrix of the 2D-DFT exhibits the centrosymmetry property, since the input matrix of the 2D-DFT, i.e., ${\textrm {Re}}(\boldsymbol\Gamma_m)$, is real-valued \cite{oppenheim2001discrete}. \textcolor{black}{Secondly, the velocities corresponding to the peaks within $\{|{\bar Y}[k,n]|,\ k=\lfloor G/2\rfloor+1,\cdots,G,$ $ n=1,\cdots,\lfloor N_\textrm{sub}/2\rfloor\}$ are unrealistically large for a typical vehicular network, according to (\ref{velocityhat}).} Due to the above two reasons, only a quarter of the elements of $|{\bf \bar Y}|$, i.e., $\{|{\bar Y}[k,n]|,\ k= 1,\cdots,\lfloor G/2\rfloor,$ $ n=1,\cdots,\lfloor N_\textrm{sub}/2\rfloor\}$,} need to be involved in the search for peaks. As a result, the computational complexity is reduced. Furthermore, due to the high-attenuation side lobes of the aliased sinc function, ${\bar Y}[k,n]$ can be approximately formulated as
			\begin{equation}
				\begin{aligned}
					&{\bar Y}[k,n]\\
					&\approx\!\frac{1}{4}\sum_{l=1}^{L}\tilde{\alpha}_l[m]{S}_{G}[(k\!-\xi_{l}\frac{N_{\textrm s}}{N_\textrm{sub}})f_{\textrm R}] 
					{S}_{N_\textrm{sub}}[( n-\frac{ \tau_l}{T_{\textrm s}})T_{\textrm R}]\!+\!\bar{{Z}}[k,\!n].
				\end{aligned}
		\end{equation}}
		
		Thus, it becomes evident that the peak coordinates $(\kappa_l,\epsilon_l)$ for the $l$th component output matrix of the 2D-DFT of ${\textrm {Re}}(\boldsymbol{\Gamma}_m)$ satisfy 
		\begin{equation}\label{equ23_1}
			\begin{cases}
				\kappa_l\!=\arg \mathop{\min}\limits_{k\in\mathbb{Z}^+}|(k-\xi_{l}\frac{N_{\textrm s}}{N_\textrm{sub}})f_{\textrm R}|^2,\\
				\epsilon_l\!=\arg \mathop{\min}\limits_{n\in\mathbb{Z}^+}|( n-\frac{ \tau_l}{T_{\textrm s}})T_{\textrm R}|^2,
			\end{cases}
		\end{equation} where $\mathbb{Z}^+$ represents the set of non-negative integers. As a result, the peak coordinates for the $l$th component output matrix of the 2D-DFT are $({\textrm {Round}}(\xi_{l}\frac{N_{\textrm s}}{N_\textrm{sub}}),$ ${\textrm {Round}}(\frac{\tau_{l}\!}{T_{\textrm s}}))$, $l=1,\cdots,L$.
		Then, the desired {range-velocity parameters} can be estimated by locating all the peaks of $|{\bf \bar Y}|$. 
		
		After determining the coordinates of the $l$th peak value as $(\kappa_l,\epsilon_l)$, $l=1,\cdots,L$, we can obtain the estimated range-velocity parameters by $\hat{r}_l=(\epsilon_l-1)R_{\textrm u}$ and $\hat{v}_l=(\kappa_l-1)V_{\textrm u}$, where $R_{\textrm u}$ and $V_{\textrm u}$ are defined as the range resolution and velocity resolution, respectively. Specifically, $R_{\textrm u}$ and $V_{\textrm u}$ are formulated as
		\begin{equation}\label{tauhat}
			R_{\textrm u} = {cT_{\textrm R}=} \frac{c}{N_\textrm{sub}\Delta f},
		\end{equation}
		\begin{equation}\label{velocityhat}
			V_{\textrm u} = {cf_{\textrm R}/f_\textrm{c}=} \frac{c}{f_\textrm{c}T_{\textrm {sym}}G}.
		\end{equation}

		According to (\ref{tauhat}) and (\ref{velocityhat}), we can see that small values of $G$ or $N_\textrm{sub}$ degrade {the corresponding resolution}. Meanwhile, large $G$ and $N_\textrm{sub}$ increase the sensing latency and computational complexity. \textcolor{black}{To improve the resolution without significantly increasing the sensing latency and computational complexity, the zero padding technique can be employed to expand the dimension of $\boldsymbol{\Gamma}_m$. As a result, the estimation performance can be enhanced. For instance, as far as the range estimation is concerned, we can apply a $K$-times zero-padding, which means $(K-1)N_\textrm{sub}$ zeros are padded, the resolution of the estimated range $\hat{r}_l$ can be improved to $c/(KN_\textrm{sub}\Delta f)$ \cite{oppenheim2001discrete}. 
			{Nevertheless, it is important to note that the effectiveness of zero padding depends on the degree of overlap of the mainlobes of the aliased sinc functions in ${\mathbf{\bar{Y}}}$.} If the mainlobes significantly overlap, zero padding may not be effective. }

		\subsection{DOA Estimation and Association}
		Various DOA estimation algorithms have been proposed in the open literature, including MUSIC \cite{schmidt1986multiple}, Estimating Signal Parameter via Rotational Invariance Techniques (ESPRIT) \cite{roy1989esprit,Hu_ESPRIT}, and the compressed sensing based DOA estimation algorithms \cite{fortunati2014single}. By exploiting any of these algorithms, one can estimate the individual DOAs ${\phi}_{l}$, $l=1,\cdots,L_\textrm{V}$. 
		
		Upon finishing the DOA estimation, the set \textcolor{black}{$\{\hat{\phi}_{i}\}_{ i=1}^{L_\textrm{V}}$} is obtained, where $\hat{\phi}_{i}$ represents the estimate of the DOA of signals propagated from the $i$th path. However, since these DOA estimates are generally obtained independently of the paired range-velocity estimation, it is challenging to associate them with the paired range-velocity parameters belonging to the \textcolor{black}{same} individual targets. To achieve the association, we propose the following association algorithm. 
		
		\textcolor{black}{Specifically, for any given $l=1,\cdots, L_{\textrm{V}}$, let us define $ {b_l} \in \{1,\cdots,L_\textrm{V}\}$ as the optimal association index between the DOA estimates $\{\hat{\phi}_{i}\}_{ i=1}^{L_\textrm{V}}$ and the $l$th path's range-velocity estimates, which are determined by the peak coordinates $(\kappa_l,\epsilon_l)$. Then, we have
			\begin{equation}\label{equ25}
				{b_l} = \arg \mathop{\max}\limits_{i\in \{1,\cdots,L_\textrm{V}\}} | {\Pi}_i[\kappa_l,\epsilon_l] |,
			\end{equation}
			where ${\Pi}_i[\kappa_l,\epsilon_l]$ is the $(\kappa_l,\epsilon_l)$th element of $\boldsymbol{\Pi}_i$ that is formulated as} 
		\begin{equation}\label{equ26}
			\boldsymbol{\Pi}_i = {\bf F}^{\textrm H}_G{\textrm {Re}}({\bf A}_i\boldsymbol{\Xi}){\bf F}^{\textrm H}_{N_\textrm{sub}}.
		\end{equation}Moreover, we have $\boldsymbol{\Xi} = [{\breve{\bf Y}^{\textrm T}_1},\cdots,{\breve{\bf Y}^{\textrm T}_G}]^{\textrm T}\in\mathbb{C}^{M_{\textrm R}G\times N_\textrm{sub}}$, and ${\bf A}_i\in\mathbb{C}^{ {G\times M_{\textrm R}G}}$ is defined as
		\begin{equation}\label{Ai}
			{\bf A}_i = {\textrm {diag}}\underbrace{({\bf a}(M_\textrm{R},\hat{\phi}_i),\cdots,{\bf a}(M_\textrm{R},\hat{\phi}_i))}_{\text{repeat}\: G \: \text{times}}.
		\end{equation}
		
		According to the definitions of $\boldsymbol{\Xi}$ and ${\bf A}_i$, we obtain 
		\begin{equation}\label{equ23}
			{\bf A}_i\boldsymbol{\Xi}=[[{\bf a}(M_\textrm{R},\hat{\phi}_i){\breve{\bf Y}_1}]^{\textrm T},\cdots,[{\bf a}(M_\textrm{R},\hat{\phi}_i){\breve{\bf Y}_G}]^{\textrm T}]^{\textrm T},
		\end{equation}
		where ${\bf A}_i$ can be regarded as a spatial filter and we have
		\begin{equation}\label{equ24}\begin{small}
				\begin{aligned}
					&{\bf a}(M_\textrm{R},\hat{\phi}_i){\breve{\bf Y}_g}=\sum_{l=1}^{L}\!\bar{h}_{l}e^{-j2\pi \xi_{l}(g-1)\frac{N_{\textrm s}}{N_\textrm{sub}}}(1\!-\!e^{j2\pi \xi_{l}G_\textrm{d}\frac{N_{\textrm s}}{N_\textrm{sub}}})  \\
					&\cdot e^{-j2\pi \xi_{l}\frac{N_\textrm{cp}}{N_\textrm{sub}}}{\bf a}(M_\textrm{R},\hat{\phi}_i){\bf a}^{\textrm T}(M_\textrm{R},\phi_{l}) {\bf a}(M_\textrm{U},\theta_{l}){\bf w}\boldsymbol{\tau}_{l}\!+\!{\bf a}(M_\textrm{R},\hat{\phi}_i)\breve{\bf Z}_g.
			\end{aligned}\end{small}
		\end{equation}
		
		Then, by performing \textcolor{black}{2D-DFT} on the filtered signal ${\textrm {Re}}({\bf A}_i\boldsymbol{\Xi})$, its delay-Doppler spectrum, \textcolor{black}{namely} $\boldsymbol{\Pi}_i$, is obtained. Obviously, $|{\Pi}_i[\kappa_l,\epsilon_l]|$ represents the square root of the power of the filtered signal components propagating from the $l$th path. Hence, it is intuitive that $|{\Pi}_i[\kappa_l,\epsilon_l]|$ will be maximized when the particular DOA $\hat{\phi}_{i}$ is the actual DOA of the $l$th path. On the contrary, if $\hat{\phi}_{i}$ is not the actual DOA corresponding to the $l$th path, $|{\Pi}_i[\kappa_l,\epsilon_l]|$ will be small, since in this case the designed spatial filter ${\bf A}_i$ will filter out the signals propagating from the $l$th path, according to (\ref{equ24}). Consequently, \textcolor{black}{$\hat \phi_{b_l}$} is associated with the paired parameters {$(\kappa_l,\epsilon_l)$ and in turn, with $(\hat{v}_l,\hat{r}_l)$}. 
		
		The proposed DOA association algorithm demonstrates robust performance. Consider an example as shown in Fig.~\ref{DOA}, regardless of the value of $M_{\textrm R}$, the magnitude of ${\bf a}(M_\textrm{R},\hat{\phi}_i){\bf a}^{\textrm T}(M_\textrm{R},\phi_{l})$ possesses a narrow lobe. Therefore, \textcolor{black}{although  $\hat \phi_i, i\neq b_l$, may be incorrectly associated with the paired parameters $(\hat{v}_l,\hat{r}_l)$, $\hat \phi_i$ will be similar to $\hat \phi_{b_l}$, thanks to the narrow lobe.}
		\begin{figure}[tbp]
			\centering
			\includegraphics[width=2.5in]{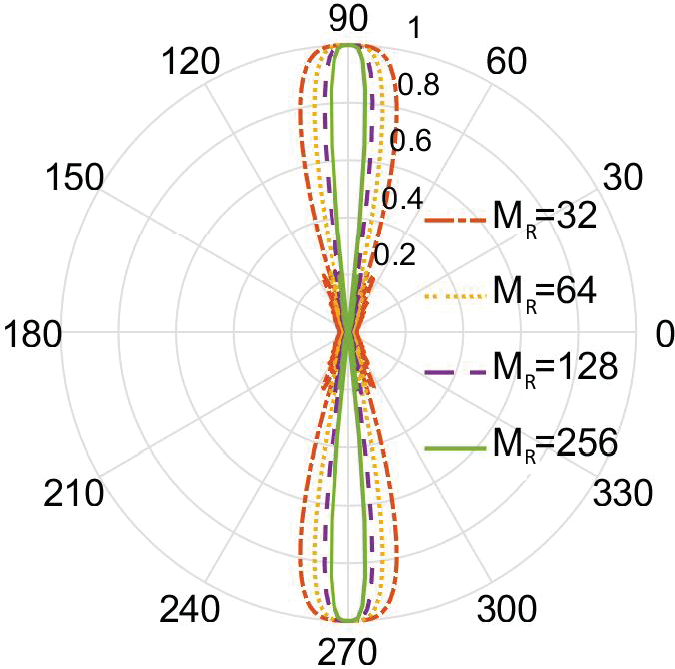}
			\caption{Magnitude of ${\bf a}(M_\textrm{R},\hat{\phi}_i){\bf a}^{\textrm T}(M_\textrm{R},\phi_{l})$  {with different values of $\hat{\phi}_i$, while $\phi_{l}$ is assumed to be $\pi/2$ in this example.}}
			\label{DOA}
		\end{figure}
		
		Moreover, the proposed DOA association algorithm can also \textcolor{black}{distinguish} the targets with similar range-velocity parameter values. To elaborate a little further, when the range-velocity parameters of different targets take highly similar values, the conventional estimation algorithms, such as the 2D-DFT based algorithm \cite{zhang2021enabling} or the MUSIC based algorithm \cite{liu2020super}, cannot identify these targets. \textcolor{black}{This is because the peaks corresponding to these targets will overlap in the delay-Doppler domain.} Fortunately, by leveraging the proposed DOA association algorithm, signal components with similar or even identical range-velocity parameter values but different \textcolor{black}{DOAs},  can be distinguished \textcolor{black}{with the aid of the designed spatial filter}.
		
		Nevertheless, the computational complexity of the association algorithm is high due to the large dimensions of  ${\bf A}_i$ and ${\bf \Xi}$. To reduce the complexity, ${\bf A}_i$ and $\boldsymbol{\Xi}$ can  {be degenerated into} ${\bf a}(M_\textrm{R},\hat{\phi}_i)$ and  {${\breve{\bf Y}}_g\in\mathbb{C}^{M_{\textrm R}\times N_\textrm{sub}}$}, respectively. Then, \eqref{equ25} can be reformulated as
		\begin{equation}\label{newequ25}
			{b_l} = \arg \mathop{\max}\limits_{i\in \{1,\cdots,L_\textrm{V}\}} |{\Pi}_i[\epsilon_l]|,
		\end{equation}
		where we have
		\begin{equation}\label{simplifiedPi}
			\boldsymbol{\Pi}_i = {\textrm {Re}}({\bf A}_i\boldsymbol{\Xi}){\bf F}^{\textrm H}_{N_\textrm{sub}}.
		\end{equation}
		Note that $\boldsymbol{\Pi}_i$ in (\ref{simplifiedPi}) has become a vector, and it represents the delay domain spectrum of the \textcolor{black}{signals ${\textrm {Re}}({\bf A}_i\boldsymbol{\Xi})$}.  As pointed out before, the physical meaning of ${\bf A}_i\boldsymbol{\Xi}$ is that \textcolor{black}{by utilizing the spatial filter ${\bf A}_i$ to filter the signals $\boldsymbol{\Xi}$, we can reserve the signals with DOA $\hat \phi_i$ and filter out signals with other DOAs}. Moreover, to ensure the signals utilized in $\boldsymbol{\Xi}$ \textcolor{black}{exhibit a high SNR, the OFDM symbol index $g$ should be appropriately selected}. 
			
			Similarly, $\boldsymbol{\Pi}_i$ in (\ref{equ26}) can also be degraded into a vector that represents the Doppler domain spectrum by defining  {$\boldsymbol{\Pi}_i={\textrm {Re}}({\bf A}_i\boldsymbol{\Xi}){\bf F}^{\textrm H}_G$}, where ${\bf A}_i$ and $\boldsymbol{\Xi}$ is approximated by ${\bf a}(M_\textrm{R},\hat{\phi}_i)$ and $[{\breve{ \bf Y}}_1[:,n],\cdots,{\breve{\bf  Y}}_G[:,n]]\in\mathbb{C}^{M_{\textrm R}\times G}$, respectively. The index $n$ should also be carefully selected to guarantee \textcolor{black}{a high SNR.} In this case, \eqref{equ25} is reformulated as
			\begin{equation}\label{newequ26}
				{b_l} = \arg \mathop{\max}\limits_{i\in \{1,\cdots,L_\textrm{V}\}} |{\Pi}_i[\kappa_l]|.
			\end{equation} 
			
			It is worth noting that regardless of which simplified algorithm is used, the association \textcolor{black}{accuracy} will be affected by the utilized window functions when conducting 2D-DFT. Numerical simulations will be carried out to further characterize the impact of different window functions. For clarity, the proposed clutter suppression and uplink passive sensing algorithm for synchronous PVNs is summarized as Algorithm~1.

			\begin{algorithm} \label{table}
				\small
				\caption{ 
					Clutter suppression and uplink passive sensing in synchronous systems} 
				\hspace*{0.02in} {\bf Input:} 
				$M_\textrm{R}, M_\textrm{U}$, $\{{\bf Y}_g\}_{g=1}^{G}$,  $\{{\bf x}_g\}_{g=1}^{G}$.
				\begin{algorithmic}[1]
					
					\State
					\textcolor{black}{ Compensate $\{{\bf Y}_g\}_{g=1}^{G}$ by \eqref{data_compensation}, and acquire $\{\breve{\bf Y}_g\}_{ g=1}^{G}$ by \eqref{equ18}.} \vspace{2pt}
					\State \textcolor{black}{Construct $\boldsymbol{\Gamma}_m$ by stacking $\breve{\mathbf{y}}_{g,m}$ as $[\breve{\mathbf{y}}_{1,m}^{\textrm T},\cdots,\breve{\mathbf{y}}_{G,m}^{\textrm T}]^{\textrm T}$, 
						and apply 2D-DFT to $\boldsymbol{\Gamma}_m$ as \eqref{2D-DFT-1}.} 
					\State Find the peak coordinates $(\kappa_l,\epsilon_l)$ by \eqref{equ23_1}, and then estimate  $\hat{r}_{l}$ and $\hat{v}_{l}$ by $\hat{r}_l =(\epsilon_l-1)R_{\textrm u}$ and $\hat{v}_{l}=(\kappa_l-1)V_{\textrm u}$, respectively, for $l=1,\cdots,L_\textrm{V}$.
					\State Estimate $\{\hat{\phi}_{i}\}_{ i=1}^{L_\textrm{V}}$ by utilizing MUSIC or any other DOA estimation algorithm. \vspace{2pt}
					\If{\ \ the full DOA association algorithm is utilized }
					\State
					\textcolor{black}{ Construct $\boldsymbol{\Xi} = [{\breve{\bf Y}^{\textrm T}_1},\cdots,{\breve{\bf Y}^{\textrm T}_g} ,\cdots, {\breve{\bf Y}^{\textrm T}_G}]^{\textrm T} $ and ${\bf A}_i$ by (\ref{Ai}), for $i=1,\cdots,L_\textrm{V}$.}
					
					\State
					\textcolor{black}{ Associate $\{\hat{\phi}_{i}\}_{ i=1}^{L_\textrm{V}}$ with their corresponding  {paired velocity-range parameters} by (\ref{equ25}).} 
					\ElsIf{ the simplified delay-domain DOA association algorithm is utilized }
					\State Construct  $\boldsymbol{\Xi}= {{\breve{\bf Y}}_g}$ and ${\bf A}_i={\bf a}(M_\textrm{R},\hat{\phi}_i)$, for $i=1,\cdots,L_\textrm{V}$. 
					\State \textcolor{black}{Associate $\{\hat{\phi}_{i}\}_{ i=1}^{L_\textrm{V}}$ with their corresponding  {paired velocity-range parameters}, which have been jointly estimated, by  (\ref{newequ25}).}\vspace{2pt}
					\Else{ the simplified Doppler-domain DOA association algorithm is utilized }
					\State Construct  
					$\boldsymbol{\Xi}= {[{\breve{\bf Y}}_1[:,n],\cdots\hspace{-2pt},\hspace{-2pt}{\breve{\bf Y}}_G[:,n]]}$ and ${\bf A}_i={\bf a}(M_\textrm{R},\hat{\phi}_i)$, for $i=1,\cdots,L_\textrm{V}$.
					\State \textcolor{black}{Associate $\{\hat{\phi}_{i}\}_{ i=1}^{L_\textrm{V}}$ with their corresponding  {paired velocity-range parameters}, which have been jointly estimated, by (\ref{newequ26}).}
					\EndIf		
				\end{algorithmic}
				\hspace*{0.02in} {\bf Output:} 
				The associated parameters: $(\hat{r}_{i}, \hat{v}_{i}, \hat{\phi}_{i})$, for $i=1,\cdots,L_\textrm{V}$.
			\end{algorithm}

			\section{ {The Cross-Multipath Cross-Correlation (CMCC) Synchronization Algorithm}}
			In  {practical} asynchronous PVNs, not only \textcolor{black}{are} the clutter suppression algorithms not directly applicable, but also the range-velocity estimation is inaccurate due to the presence of CFO and TO \cite{ni2021uplink}. Therefore, there is a critical need for high-performance CFO and TO \textcolor{black}{estimation and synchronization algorithms}. Although  {several} time-frequency synchronization algorithms have been introduced to align transceivers, they are predominantly tailored for the LOS environment and/or the {multi-antenna-receiver} scenario, and  {are} not applicable to NLOS or single-antenna-receiver scenarios. In response to this challenge, we propose a novel synchronization algorithm applicable  {to multi-target sensing in \textcolor{black}{both} the single-antenna-receiver/multi-antenna-receiver and the LOS/NLOS scenarios.}  {The role of the proposed synchronization algorithm in the whole sensing framework is depicted in Fig. \ref{fig2}.}

			Before introducing the design idea of the proposed synchronization algorithm,  a common \textcolor{black}{synchronization framework utilized by existing ISAC algorithms, e.g., \cite{IndoTrack,ni2021uplink,FarSense}}, is summarized as follows. This framework involves two key steps: 1) initially, these algorithms select the signal components received from a particular antenna as their synchronization reference \textcolor{black}{signal};
			2) subsequently, they perform a cross-correlation operation \cite{IndoTrack,ni2021uplink}, or a simple division operation \cite{FarSense}, between the reference signal \textcolor{black}{and the signals received by other antennas} to extract CFO and TO. 
			However, this prevailing synchronization framework falls short in addressing the challenges present in the NLOS and the single-antenna-receiver scenarios. 
			
			To overcome this limitation, we design a synchronization algorithm based on an innovative framework.  {Concretely, in} our proposed algorithm, we designate the multipath signals reflected \textcolor{black}{from static objects}, commonly \textcolor{black}{referred as} ``clutter", to be the synchronization reference signal.  {This kind of signals carry the location information of the static \textcolor{black}{objects.} As a result, \textcolor{black}{they} can be deemed as a kind of ``identification sequence" of the static \textcolor{black}{environment}. Since the locations of the static objects generally remain unchanged during a long period of time, the ``identification sequence" of the static environment can also be regarded as constant. However, the CFO and TO vary in a shorter period and the change of CFO and TO may be reflected in the ``identification sequence" of the whole environment. In light of this, we intend to investigate how the ``identification sequence" evolves in response to the changes in CFO and TO. Upon determining the pattern of changes, it becomes feasible to deduce the specific values of CFO and TO by monitoring variations in the ``identification sequence".}

			To facilitate the exposition of the above idea, we firstly reconsider the formulation of \textcolor{black}{${ \bar Y}[k,n]$}, as presented in (\ref{delayspectrum}), in the context of asynchronous systems to characterize the frequency spectrum of the clutter. As pointed out in Section \ref{sec:clutter_suppression}, we have $\xi_{l}=\xi_{{\textrm D},l}+\xi_{\textrm{o}}$ \textcolor{black}{and  $\tau_{l}=\tau_{{\textrm d},l}+\tau_{\textrm{o}}$}. Furthermore, since the CIR associated with the clutter exhibits a \textcolor{black}{near-zero} Doppler frequency shift according to (\ref{clutterchannel}), \textcolor{black}{the frequency spectrum} corresponding to the clutter is centred at $\frac{\xi_{\textrm{o}} N_{\textrm s}K}{N_\textrm{sub}}f_{\textrm R}$ \textcolor{black}{according to (\ref{equ23_1}).}
			Then, we can select the $K_{\textrm c}={\textrm {Round}}(\frac{\xi_{\textrm{o}} N_{\textrm s}K}{N_\textrm{sub}})$th row of ${\bf \bar Y}$ to be the ``identification sequence". In the following, we \textcolor{black}{name} the sequence as the fingerprint spectrum for clarity.
			\textcolor{black}{As previously described,} the row, corresponding to the delay-Doppler domain spectrum of clutter, is composed of signals reflected by static objects. Obviously, $\xi_{l}$  \textcolor{black}{equals $\xi_\textrm{o}$ for all the clutter paths. 
				Hence,} we have
			\begin{equation}
				\begin{aligned}
					&{\bar Y}[K_\textrm{c},n] \\
					&\!\approx\! \frac{1}{4}\sum_{l=1}^{L}\tilde{\alpha}_l[m]{S}_{G}[(K_\textrm{c}\!-\!\xi_{l}\frac{N_{\textrm s}}{N_\textrm{sub}})f_{\textrm R}] 
					{S}_{N_\textrm{sub}}[(n\!-\!\frac{ \tau_l}{T_{\textrm s}})T_{\textrm R}] \!+\!\bar{{Z}}[K_\textrm{c},\!n]\\
					&\!\approx\!\frac{1}{4}{S}_{G}[(K_\textrm{c}\!-\!\xi_{\textrm o}\frac{N_{\textrm s}}{N_\textrm{sub}})f_{\textrm R}]\sum_{l=1}^{L}\tilde{\alpha}_l[m] 
					{S}_{N_\textrm{sub}}[( n\!-\!\frac{ \tau_l}{T_{\textrm s}})T_{\textrm R}]\!+\!\bar{{Z}}[K_\textrm{c},\!n].
				\end{aligned}
			\end{equation} 
			Then, upon assuming that the CFO $\xi_\textrm{o}$ and TO $\tau_\textrm{o}$ increase by $\Delta\xi_{\textrm{o}}$ and $\Delta\tau_\textrm{o}$, respectively, the $n$th element of the \textcolor{black}{updated} fingerprint spectrum ${\bar Y}_{\textrm {up}}[K_{\textrm c},n]$ can be formulated as 
			\begin{equation}\label{newdelayspectrum2}\begin{small}
					\begin{aligned}
						&{\bar Y}_{\textrm {up}}[K_{\textrm c},n]\approx\\
						&\frac{1}{4}{S}_{G}([K_{\textrm c}-\xi_{\textrm{o}}\frac{N_{\textrm s}K}{N_\textrm{sub}}-{\textrm {Int}}(\Delta\xi_{\textrm{o}}\frac{N_{\textrm s}K}{N_\textrm{sub}})-{\textrm {Frac}}(\Delta\xi_{\textrm{o}}\frac{N_{\textrm s}K}{N_\textrm{sub}})]f_{\textrm R})
						\sum_{l=1}^{L}\\
						&\tilde{\alpha}_l[m] 
						{S}_{N_\textrm{sub}}([n- \frac{(\tau_\textrm{o}+\tau_{{\textrm d},l})K}{T_{\textrm s}}-{\textrm {Int}}(\Delta\tau_\textrm{o}\frac{K}{T_{\textrm s}})-{\textrm {Frac}}(\Delta\tau_\textrm{o}\frac{K}{T_{\textrm s}})]T_{\textrm R})\\
						& +\bar{{Z}}_{\textrm{up}}[K_{\textrm c},n]\approx{\bar Y}\left[K_{\textrm c}-{\textrm {Round}}(\Delta\xi_{\textrm{o}}\frac{N_{\textrm s}K}{N_\textrm{sub}}),n-{\textrm {Round}}(\Delta\tau_\textrm{o}\frac{K}{T_{\textrm s}})\right].
				\end{aligned}\end{small}
			\end{equation}
			\textcolor{black}{The above relationship} holds when the location distributions of the background objects in the environment and the UT \textcolor{black}{remain} unchanged or move slowly.\footnote{In practice, the displacement of the UT and the background objects \textcolor{black}{is} typically small enough to be neglected on a time-scale of tens of milliseconds, or even seconds. Moreover, we can select a static UT as the transmitter of the sensing signals, as done in \cite{ni2021uplink,FarSense,zeng2020multisense}. However, with the CFO and the TO drifting constantly, $\Delta\xi_{\textrm{o}}$ and $\Delta\tau_\textrm{o}$ can be large on the time-scale of seconds, or even tens of milliseconds.}  
			Note that with the $K$-times zero-padding, the frequency resolution and time resolution introduced in (\ref{DFT1}) and (\ref{DFT2}), i.e., $f_{\textrm R}$ and $T_{\textrm R}$, become $\frac{\Delta f}{KG}$ and $\frac{T_\textrm{s}}{K}$, respectively. 
			
			Note that the selected zero-padding coefficient $K$ is generally large for achieving high frequency resolution and time resolution, thus resulting in sufficiently small ${\textrm {Frac}}(\Delta\xi_{\textrm{o}}\frac{N_{\textrm s}K}{N_\textrm{sub}})$ and ${\textrm {Frac}}(\Delta\tau_\textrm{o}\frac{K}{T_{\textrm s}})$.  Then, we have ${\textrm {Frac}}(\Delta\xi_{\textrm{o}}\frac{N_{\textrm s}K}{N_\textrm{sub}})f_{\textrm R}<\frac{\Delta f}{KG}$ and ${\textrm {Frac}}(\Delta\tau_\textrm{o}\frac{K}{T_{\textrm s}})T_{\textrm R}<\frac{T_\textrm{s}}{K}$. Thus, the small impact of ${\textrm {Frac}}(\Delta\xi_{\textrm{o}}\frac{N_{\textrm s}K}{N_\textrm{sub}})$ and ${\textrm {Frac}}(\Delta\tau_\textrm{o}\frac{\!K}{T_{\textrm s}})$ in (\ref{newdelayspectrum2}) can be neglected with large $K$, and ${\bf \bar Y}_{\textrm {up}}$ can be deemed as a circularly shifted version of ${\bf \bar Y}$.
			
			So far, we have drawn the conclusion that {the fingerprint spectrum approximately exhibits a  circular shift property in response to the changes of CFO and TO. Specifically, if CFO and TO increase by $\Delta\xi_{\textrm{o}}$ and $\Delta\tau_\textrm{o}$, respectively, the fingerprint spectrum will circularly shift ${\textrm {Round}}\left(\Delta\xi_{\textrm{o}}\frac{N_{\textrm s}K}{N_\textrm{sub}}\right)$ and ${\textrm {Round}}\left(\Delta\tau_\textrm{o}\frac{K}{T_{\textrm s}}\right)$ respectively, in the direction of increasing the Doppler frequency shift and the delay.}

			By exploiting the change pattern between the fingerprint spectrum and the CFO/TO, the CFO/TO estimation algorithm is proposed. To begin with, we denote ${\bf \bar Y}[K_{\textrm c},:]$ as the  {\textit{original fingerprint spectrum}, ${\boldsymbol \zeta}$, in what follows for convenience. Then, since the offset fingerprint spectrum is the circular shift of the original fingerprint spectrum,} cross-correlation between ${\boldsymbol \zeta}$ and ${\bf \bar Y}_{\textrm {up}}$ can be performed to estimate the CFO and TO as 
			\begin{equation}\label{unsimplified}\small
				\begin{aligned}
					\{\!\Delta\hat\xi_{\textrm{o}}\frac{N_{\textrm s}K}{N_\textrm{sub}}, \Delta\hat\tau_\textrm{o} \frac{K}{T_{\textrm s}}\!\}\!\!=\arg \mathop{\max}\limits_{k,q}\bigg|\!\sum_{i=1}^{Q_\textrm{B}}{\bar Y}_{\textrm {up}}[k,(q+i)\,{\textrm {mod}}\,Q_\textrm{B}]{\boldsymbol \zeta}^*(i)\bigg|,
				\end{aligned}\small
			\end{equation}
			where we have $k=1,\cdots,  K_\textrm{B} $ and $q=1,\cdots, Q_\textrm{B} $, with $K_\textrm{B} = \lfloor\frac{KG}{2}\rfloor$ and $Q_\textrm{B} = \lfloor\frac{KN_\textrm{sub}}{2}\rfloor$. 
			As a result, by using (\ref{unsimplified}),  $\Delta\tau_\textrm{o}\frac{K}{T_{\textrm s}}$ and $\Delta\xi_{\textrm{o}}\frac{N_{\textrm s}K}{N_\textrm{sub}}$ can be estimated. 
			However, the computational complexity of conducting the 2-D maximum likelihood (ML) search \textcolor{black}{of} (\ref{unsimplified}) is high. To reduce the computational complexity, a simplified version of CMCC,  {called simplified CMCC (S-CMCC)}, is  {designed} by decomposing the 2-D ML search into two separate 1-D ML searches.
			
			The first 1-D ML search is dedicated to the CFO estimation. Specifically, we aim to estimate the CFO  {by locating the row of $\bar{\bf Y}_{\textrm {up}}$, which has the closest power with that of} the original fingerprint spectrum. According to Parseval's Theorem, the power of the original fingerprint spectrum is equal to the power of the signals reflected from static objects,  while the power of this kind of signals is determined by the factors, such as the radar cross section (RCS) of  static objects and the transmitting power of UT\footnote{Actually, these factors take approximately constant values on the time-scale of hundreds of milliseconds or even seconds.}. Therefore, if the location \textcolor{black}{distributions} of the  static objects and the UT, as well as the transmitting power of the UT, remain unchanged, the power of the offset fingerprint spectrum, will be approximately equal to that of the original fingerprint spectrum. As a result, the 1-D ML search can be formulated as
			\begin{equation}\label{hatg}
				\Delta\hat\xi_{\textrm{o}} = \frac{N_\textrm{sub}}{KN_{\textrm s}}\arg \mathop{\min}\limits_{k}\big| |{\boldsymbol \zeta}|^2-\sum_{n=0}^{Q_\textrm{B}}|{\bar Y_\textrm{up}}[k, {n\ {\textrm {mod}}\ Q_\textrm{B}}]|^2 \big|.
			\end{equation}
			
			Upon completing the estimation of $\Delta\xi_{\textrm{o}}$, another  {1-D ML} search can be performed to estimate  {$\Delta\tau_\textrm{o}$}. The estimate is given by
			\begin{equation}\label{hatn}
				\Delta\hat\tau_\textrm{o}\!=\! \frac{T_{\textrm s}}{K}\arg \mathop{\max}\limits_{q}  \bigg|\sum_{i=1}^{Q_\textrm{B}}{\boldsymbol{\zeta}^*(i)\bar{Y}_\textrm{up}[ {\Delta\hat\xi_{\textrm{o}}\frac{N_{\textrm s}K}{N_\textrm{sub}}},(i\!+\!q)\ {\textrm {mod}} \ Q_\textrm{B}]} \bigg|.
				\small
			\end{equation}
			In summary, the proposed synchronization algorithm is {presented} in Algorithm\ 2.
			\begin{algorithm}
				\small
				
				\caption{The proposed synchronization algorithm} 
				\hspace*{0.02in} {\bf Input:} 
				$\bar{\bf Y}$.	
				\begin{algorithmic}[1]
					\If{\ \ 	\textcolor{black}{using CMCC}}
					\State Find the exact CFO $\Delta\hat\xi_{\textrm{o}}$ and TO $\Delta\hat\tau_\textrm{o}$ by correlating the \textcolor{black}{updated} fingerprint spectrum and the original fingerprint spectrum according to \eqref{unsimplified}.
					\Else{\ \ using S-CMCC} {\bf then}
					\State Find the exact CFO $\Delta\hat\xi_{\textrm{o}}$ by \eqref{hatg}. 
					\State Find the exact TO $\Delta\hat\tau_\textrm{o}$ by correlating the \textcolor{black}{updated} fingerprint spectrum and the original fingerprint spectrum according to \eqref{hatn}.		
					\EndIf
				\end{algorithmic}
				\hspace*{0.02in} {\bf Output:} 
				$\Delta\hat\xi_{\textrm{o}}, \Delta\hat\tau_\textrm{o}$.
			\end{algorithm}
			\section{Performance Analysis}
			In this section, we begin by deriving the CRLB for the estimated paired range-velocity parameters upon implementing the clutter suppression.  Then, we derive the MSE of the proposed synchronization algorithm.
			
			\subsection{CRLB of the Range-Velocity Estimate}
			We first reconstruct the matrix $\boldsymbol{\Gamma}_m$ in (\ref{Gamma}) as
			\begin{equation}\label{appendixb2}
				\begin{aligned}
					\boldsymbol{\Gamma}_m 
					=\sum_{l=1}^{L}\beta_l\boldsymbol{\xi}_l\boldsymbol{\tau}_l+\tilde{\bf Z},
				\end{aligned}
			\end{equation}
			where $\beta_l$ is formulated as
			\begin{equation}\label{beta_l}
				\beta_l \!=\! {{\bar h}_{l}}e^{\frac{-j2\pi (m\!-\!1)d}{\lambda}\!\sin\phi_{l}}{\bf a}(M_\textrm{U},\theta_{l}){\bf w}.	
			\end{equation}
			
			Moreover, according to \cite{stoica1990performance,stoica1989music}, the $(i,j)$th element of the CRLB matrix ${\bf J}$ can be formulated as
			\begin{equation}\label{CRLB}\small
				\begin{aligned}
					{\textrm{Re}}\left\{\frac{1}{\tilde{\sigma}^2} \sum_{n=1}^{N_\textrm{sub}}\left[\frac{\partial}{\partial {\alpha}[i]} \sum_{l=1}^{L}\beta_l\boldsymbol{\xi}_l{\tau}_l[n]\right]^{\textrm H} \!
					\left[\frac{\partial}{\partial {\alpha}[j]} \sum_{l=1}^{L}\beta_l\boldsymbol{\xi}_l{\tau}_l[n]\right]\right\}^{-1},
				\end{aligned}\small
			\end{equation} 
			where ${\alpha}[i]$ is the $i$th element of
			$\boldsymbol{\alpha}=[\xi_1,\cdots,\xi_L,\tau_1,\cdots,\tau_L]$ and $\tilde{\sigma}^2$ is defined as the power of the AWGN noise in ${\tilde {\bf Z}}$.
			
			Furthermore, we introduce the matrix ${\bf H}_n$, whose $i$th column is defined as $\frac{\partial}{\partial {\alpha}[i]} \sum_{l=1}^{L}\beta_l\boldsymbol{\xi}_l{\tau}_l[n]$. With this definition, the CRLB matrix ${\bf J}$ can be formulated as follows:
			\begin{equation}\label{Omega}
				{\bf J} = \frac{1}{\tilde{\sigma}^2}{\textrm{Re}}\left\{\sum_{n=1}^{N_\textrm{sub}} {\bf H}_n^{\textrm H}{\bf H}_n\right\}^{-1}.
			\end{equation}
			Then, the partial derivative of $\sum_{l=1}^{L}\beta_l\boldsymbol{\xi}_l{\tau}_l[n]$ with respect to $\xi_{l}$ can be formulated as
			\begin{equation} \label{partialeta}
				\frac{\partial}{\partial \xi_{l}} \sum_{l=1}^{L}\beta_l\boldsymbol{\xi}_l{\tau}_l[n]=\beta_l{\tau}_l[n]\frac{\partial \boldsymbol{\xi}_l}{\partial \xi_{l}},
			\end{equation}
			where the $g$th element of $\breve{\boldsymbol{\xi}}_l=\frac{\partial \boldsymbol{\xi}_l}{\partial \xi_{l}}$ is expressed as 
			\begin{equation}\label{partialeta1}
				\begin{footnotesize}
					\begin{aligned}
						&\breve{\xi}_l[g]=-j2\pi e^{-j2\pi \xi_{l}[(G\!-1)\frac{N_{\textrm s}}{N_\textrm{sub}}\!+\!\frac{N_\textrm{cp}}{N_\textrm{sub}}]}\left[ (G-1)\frac{N_{\textrm s}}{N_\textrm{sub}}\!+\!\frac{N_\textrm{cp}}{N_\textrm{sub}} \right]\\
						&+\!j2\pi[(G\!-\!1\!-\!G_\textrm{d})\frac{N_{\textrm s}}{N_\textrm{sub}}\!+\!\frac{N_\textrm{cp}}{N_\textrm{sub}}]e^{j2\pi\xi_{l}\frac{G_\textrm{d}N_\textrm{s}}{N_\textrm{sub}}}e^{-j2\pi \xi_{l}[(g-1)\frac{N_{\textrm s}}{N_\textrm{sub}}\!+\!\frac{N_\textrm{cp}}{N_\textrm{sub}}]}.
				\end{aligned}\end{footnotesize}
			\end{equation}The partial derivative of $\sum_{l=1}^{L}\beta_l\boldsymbol{\xi}_l{\tau}_l[n]$ with respect to ${\tau_{l}}$ is formulated as:
			\begin{equation}\label{partialtau}
				{\frac{\partial}{\partial \tau_{l}}\sum_{l=1}^{ {L}}\beta_l\boldsymbol{\xi}_l{\tau}_l[n]=-j2\pi n\Delta fe^{-j2\pi n\Delta f\tau_{l}}\beta_l\boldsymbol{\xi}_l.}
			\end{equation}
			As a result, the computation of the matrix  {${\bf H}_n=[\beta_1{\tau}_1[n]$ $\frac{\partial \boldsymbol{\xi}_1}{\partial \xi_{1}},$ $\cdots,$ $\beta_L{\tau}_L[n]\frac{\partial \boldsymbol{\xi}_L}{\partial \xi_{L}}$$,-j2\pi n\Delta f$$e^{-j2\pi n\Delta f\tau_{1}}\beta_1\boldsymbol{\xi}_1,$ $\cdots,-j2\pi n\Delta fe^{-j2\pi n\Delta f\tau_{L}}\beta_L\boldsymbol{\xi}_L]$} and ${\bf J}$ can be performed using numerical methods with the help of Equations (\ref{Omega}), (\ref{partialeta}), and (\ref{partialtau}).

			According to (\ref{partialeta1}) and (\ref{partialtau}), $\frac{\partial \boldsymbol{\xi}_l}{\partial \xi_{l}}$ and $\frac{\partial}{\partial \tau_{l}}\beta_l\boldsymbol{\xi}_l{\tau}_l[n]$, corresponding to the MTI based joint clutter suppression and sensing scheme, are different from those in the traditional sensing schemes without clutter suppression. Specifically, there is a new and complex-valued term, $j2\pi[(g-1-G_\textrm{d})\frac{N_{\textrm s}}{N_\textrm{sub}}\!+\!\frac{N_\textrm{cp}}{N_\textrm{sub}}]e^{j2\pi\xi_{l}\frac{G_\textrm{d}N_\textrm{s}}{N_\textrm{sub}}}e^{-j2\pi \xi_{l}[(g-1)\frac{N_{\textrm s}}{N_\textrm{sub}}\!+\!\frac{N_\textrm{cp}}{N_\textrm{sub}}]}$, in the $g$th element of $\frac{\partial \boldsymbol{\xi}_l}{\partial \xi_{l}}$,  compared with the sensing schemes without clutter suppression. It is difficult to determine if the CRLB of the MTI based sensing scheme is capable of approaching the CRLB of the sensing schemes without clutter suppression by directly deriving (\ref{Omega}).
			
			To analyse the reachability, we reconsider the signal model (\ref{Gamma}) in the MTI based sensing scheme. As (\ref{Gamma}) suggests,
			when $\xi_l$ and the selected $G_\textrm{d}$ satisfy $1\!-\!e^{j2\pi \xi_{l}G_\textrm{d}\frac{N_{\textrm s}}{N_\textrm{sub}}}=2$ for any $l=1,\cdots,L$, it is intuitive that the power of the signals will be four times of the signal power in traditional sensing schemes. Meanwhile, it is obvious that the noise power in the MTI based sensing scheme is two times of the noise power in traditional sensing schemes. This is because the minus operation in MTI doubles the noise power. In this case, the SNR in the MTI based sensing scheme is two times of that in the traditional sensing schemes. However, the number of available OFDM symbols for perception, after differential operations, will be halved. Consequently, with doubled SNR and the halved number of OFDM symbols, the CRLB of the MTI based sensing scheme is the same as that of the conventional sensing schemes \cite{stoica1990performance}.

			\subsection{Theoretical Synchronization MSE}
			By denoting the original fingerprint spectrum and the offset fingerprint spectrum as $\boldsymbol{\zeta}$ and ${\bf r}$ in this subsection, we can {simply} reformulate the $n$th element of  $\boldsymbol{\zeta}$ and ${\bf r}$ as 
			\begin{equation}
				\begin{aligned}
					&{\zeta}[n]={ s}_1[n]+ \check{ z}_1[n],\\ &{ r}[n] = { s}_2[n] + \check{z}_2[n],
				\end{aligned}
			\end{equation}where $\check{z}_1[n]$ and $\check{z}_2[n]$ are complex-valued AWGNs with zero mean and variances $\check{\sigma}^2_1$ and $\check{\sigma}^2_2$, respectively. Moreover,  {${\bf  s}_1={\mathrm E}({ \bf \bar Y}[K_{\textrm c},:])$ and ${\bf  s}_2={\mathrm E}({ \bf \bar Y}_{\textrm {up}}[K_{\textrm c},:])$.}

			To evaluate the performance of the proposed synchronization algorithm, we firstly  {formulate} the cross-correlation shown in (\ref{unsimplified}) between $\boldsymbol{\zeta}$ and ${\bf r}$ as \begin{equation}\label{crosscorrelation}
				\begin{aligned}
					{\varrho}_r[q] 
					\!=&\big\vert\sum_{n=1}^{Q_\textrm{B}}\!{ s}_1[n]{ s}^*_2[n\oplus q]\!+\!{ s}_1[n]\check{ z}_2^*[n\oplus q]\!\\
					&+\!{ s}_2^*[n\oplus q]\check{ z}_1[n]\!+\!\check{ z}_1[n]\check{ z}_2^*[n\oplus q]\big\vert.
				\end{aligned}
			\end{equation}where $q=1,\cdots, Q_\textrm{B} $. Around the peak of the correlation sequence, we can decompose (\ref{crosscorrelation}) into in-phase and quadrature components relative to the ${s}_1[n]{ s}_2^*[n\oplus q]$ product. Moreover, for a usable SNR, the noise power in the quadrature part can be neglected  according to \cite{650240}, leading to
			\begin{equation}\label{rhorq}
				\begin{aligned}
					{\varrho}_r[q]\approx& {\varrho}_s[q]\!+\!\sum_{n=1}^{Q_\textrm{B}}\textrm{inPhase}\{{ s}_1[n]\check{z}_2^*[n\oplus q]\\
					&+{ s}_2^*[n\oplus q]\check{ z}_1[n]+\check{z}_1[n]\check{z}_2^*[n\oplus q]\},
				\end{aligned}
			\end{equation}where ${\varrho}_s[q]$ is defined as $\sum_{n=1}^{Q_\textrm{B}}\!{ s}_1[n]{s}_2^*[n\oplus q]$, and $\textrm{inPhase}\{\cdot\}$  represents the operator that reserves signal components in the direction indicated by $\{\cdot\}$.  For brevity, we denote the second term on the right-hand-side of (\ref{rhorq}), as $\check{ z}_s[q]$. 
			Then, the vector ${\boldsymbol\varrho}_r$, with ${\varrho}_r[q]$ being its $q$th element, is  approximately a Gaussian random vector with the covariance matrix $[{\bf s}_1{\bf s}_2^{\textrm T}(\check{\sigma}_1^2+\check{\sigma}_2^2)+Q_\textrm{B}\check{\sigma}_1^2\check{\sigma}_2^2]{\bf I}_{Q_\textrm{B}}$ and the mean vector ${\boldsymbol\varrho}_s$, which is given by   
			\begin{equation}\label{newcorrelation}
				\begin{footnotesize}
					\begin{aligned}
						&{\varrho}_s[q] = \frac{1}{16}\Upsilon\sum_{l^{'}=1}^{L}\sum_{l=1}^{L}\tilde{{\alpha}}_l[m]\tilde{{\alpha}}_{l^{'}}[m]\sum_{n=1}^{Q_\textrm{B}}\\
						&{S}_{N_\textrm{sub}}([ n+\!q\!-\!\frac{(\tau_\textrm{o}\!+\!\tau_{{\textrm d},l})K}{T_{\textrm s}}]T_{\textrm R}){S}_{N_\textrm{sub}}([n\!-\!\frac{(\Delta\tau_\textrm{o}\!+\!\tau_\textrm{o}\!+\!\tau_{{\textrm d}, l^{'}})K}{T_{\textrm s}}]T_{\textrm R}),
				\end{aligned}\end{footnotesize}
			\end{equation}
			where $\Upsilon$ is defined as ${S}_{G}([K_{\textrm c}-\xi_{\textrm{o}}\frac{N_{\textrm s}K}{N_\textrm{sub}}]f_{\textrm R}){S}_{G}([K_{\textrm c}-\xi_{\textrm{o}}\frac{N_{\textrm s}K}{N_\textrm{sub}}-{\textrm {Round}}(\Delta\xi_{\textrm{o}}\frac{N_{\textrm s}K}{N_\textrm{sub}})]f_{\textrm R})$.
			
			According to (\ref{unsimplified}) and (\ref{hatn}), if the $q_0$th element of ${\boldsymbol\varrho}_r$ is the largest among all the elements of ${\boldsymbol\varrho}_r$, then the estimated TO, $\Delta\hat\tau_\textrm{o}$, will be determined as $q_0T_{\textrm s}/K$. 
			To compute the probability that ${\varrho}_r[q_0]$ is the largest among all the elements of ${\boldsymbol\varrho}_r$, we define a new Gaussian random vector ${\boldsymbol\varrho}_q = {\boldsymbol\varrho}_{r,\bar q}-\textrm{vec}({\varrho}_r[q])$, 
			where ${\boldsymbol\varrho}_{r,\bar q}$ is defined as the vector that removes the $q$th element of ${\boldsymbol\varrho}_r$ and $\textrm{vec}({\varrho}_r[q])$ is a vector having the same size as ${\boldsymbol\varrho}_{r,\bar q}$ and all of its elements are ${\varrho}_r[q]$. Note that the number of elements of ${\boldsymbol\varrho}_{r,\bar q}$ is less than that of  ${\boldsymbol\varrho}_r$ by one.
			Then, the probability that ${\varrho}_r[q_0]$ is the maximum among all the elements $\{{\varrho}_r[q]\}_{q=1}^{Q_\textrm{B}}$ can be expressed as the cumulative distribution function (CDF) of the Gaussian vector ${\boldsymbol\varrho}_{q_0}$, satisfying 
			\begin{equation}\label{pro}
				\begin{aligned}
					 {\chi}_{q_0} &= {\textrm P} \left\{{\varrho}_r[q_0]=\max \{{\varrho}_r[1],\cdots,{\varrho}_r[Q_\textrm{B}]\} \right\}\\
					&={\textrm P}\left\{{\varrho}_{q_0}[i]<0 \mid \forall i=1, \ldots, Q_\textrm{B}-1\right\},
			\end{aligned}\end{equation}where ${\textrm P}\{\cdot\}$ represents the probability of a certain event.  As a result, the MSE of the proposed  synchronization algorithm can be formulated as
			\begin{equation}\label{MSE}
				{\textrm {MSE}} = \sum_{q=1}^{Q_\textrm{B}}{\chi}_{{q}}(\frac{qT_{\textrm s}}{K}-\tau_\textrm{o})^2.
		\end{equation}}
		\ \  {However, ${\chi}_{q}$ is not in closed form, hence its value can only be obtained by performing numerical calculations. The numerical calculations impose extremely high computational complexity due to the large values of $K$ and $N_\textrm{sub}$. Therefore, in the numerical calculations, we only ensure that ${\varrho}_r[q]<{\varrho}_r[q_0]$ holds when $q$ is around $q_0$, e.g., $q\in(q_0-\nu,q_0+\nu)\cap \mathbb{Z}$,} where $\nu$ is a selected positive integer.
		
		To sum up, the MSE described in (\ref{MSE}) is related to the Gaussian vector ${\boldsymbol\varrho}_r$, which is determined by the mean ${\boldsymbol\varrho}_s$ and the AWGN ${\bf\check{z}}_s$ composed of $\check{ z}_s[q]$. The noise power of ${\bf\check{z}}_s$ is determined by the SNR of the received signals, while the mean ${\boldsymbol\varrho}_s$ is severely affected by the number and locations of the static objects in the background. However, since $\chi_{q_0}$ and $\chi_q$ in (\ref{pro}) and (\ref{MSE}) are not in closed form, it is challenging to quantitatively analyse the relationship between the distribution/the number of background objects and the MSE. Therefore, we will preliminarily evaluate the impact of the distribution and the number of background objects through numerical simulations in Section \uppercase\expandafter{\romannumeral6}.

		\subsection{Computational Complexity}
		The computational complexity of the proposed scheme in synchronous systems, as depicted in Algorithm 1, can be analysed as follows. In Step 1, the computational complexity is $\mathcal{O}((N_\textrm{sub}^2+N_\textrm{sub})M_{\textrm R}G)$. From Steps 2 to 3, the computational cost is $\mathcal{O}(\frac{1}{16}(G^2N_\textrm{sub}+4N_\textrm{sub}^2G+4L_\textrm{V}N_\textrm{sub}G ))$ according to (14). In Step 4, the computational complexity of the DOA estimation using MUSIC is $\mathcal{O}(2M_{\textrm R}^3+M_{\textrm R}^2N_\textrm{sub}+N_\textrm{A}(M_{\textrm R}^2+M_{\textrm R}))$, where $N_\textrm{A}$ is defined as the number of angles searched. From Steps 6-7, the computational complexity is {$\mathcal{O}(4G^2M_{\textrm R}N_\textrm{sub}+2G^2N_\textrm{sub}+2GN_\textrm{sub}^2)L_\textrm{V}$}. Moreover, when the simplified DOA association algorithms are utilized, either Steps 9-10 or 12-13 will be valid. Specifically, the computational complexity of Steps 9-10 can be represented as $\mathcal{O}(4M_{\textrm R}N_\textrm{sub}+2N_\textrm{sub}^2+N_\textrm{sub})L_\textrm{V}$, and that of Steps 12-13 is given by $\mathcal{O}(4M_{\textrm R}G+2G^2+G)L_\textrm{V}$.  
		
		For the synchronization algorithm presented in Algorithm 2, the computational complexity of the unsimplified synchronization algorithm (corresponds to Step 2) is $\mathcal{O}( {K^3N^2_\textrm{ sub}G/2})$, while the computational complexity of the simplified synchronization algorithm (corresponds to Steps 4 and 5) is $\mathcal{O}( {K^2GN_\textrm{sub}/2+K^2GN_\textrm{sub}^2})$.
		
		As a result, depending on specific configurations, the overall computational complexity of our proposed scheme is summarized in Table \ref{table3}.
		\begin{table*}[tbp]	\newcommand{\tabincell}[2]{\begin{tabular}{@{}#1@{}}#2\end{tabular}}  
			\centering
			\caption{The computational complexity of our scheme with different configurations}
			\renewcommand{\arraystretch}{1.2}  
			\vspace{0.3cm}
			\scalebox{1}{\begin{tabular}{c|c|c|c}  		\hline	
					\textbf{Scheme} & \textbf{Full DOA association} & \textbf{Delay-domain association} & \textbf{Doppler-domain association} \\ \hline
					\tabincell{c}{CMCC} & \tabincell{c}{$L_\textrm{V}(2GN_\textrm{sub}^2+2G^2N_\textrm{sub}+4G^2M_{\textrm R}N_\textrm{sub})$\\ $+(GK^3N_\textrm{sub}^2)/2$} & \tabincell{c}{$L_\textrm{V}(N_\textrm{sub}+2N_\textrm{sub}^2+4M_{\textrm R}N_\textrm{sub})$\\ $+(GK^3N_\textrm{sub}^2)/2$}& \tabincell{c}{$L_\textrm{V}(G+2G^2+4GM_{\textrm R})$\\ $+(GK^3N_\textrm{sub}^2)/2$}\\
					\hline
					\tabincell{c}{S-CMCC} & \tabincell{c}{$L_\textrm{V}(2GN_\textrm{sub}^2+2G^2N_\textrm{sub}+4G^2M_{\textrm R}N_\textrm{sub})$\\ $+(GK^2N_\textrm{sub})/2+GK^2N_\textrm{sub}^2$} & \tabincell{c}{$L_\textrm{V}(N_\textrm{sub}+2N_\textrm{sub}^2+4M_{\textrm R}N_\textrm{sub})$\\ $+(GK^2N_\textrm{sub})/2+GK^2N_\textrm{sub}^2$} & \tabincell{c}{$L_\textrm{V}(G+2G^2+4GM_{\textrm R})$\\ $+GK^2N_\textrm{sub}^2+(GK^2N_\textrm{sub})/2$}\\
					\hline
			\end{tabular}}
			\label{table3}
		\end{table*}

		\section{Simulations and Discussions}
		In this section we perform Monte Carlo simulations to evaluate the performance of our proposed algorithms. In the simulations, we assume that the number of antenna-elements on each RRU and each UT is $M_\textrm{R}=64$ and $M_\textrm{U}=2$, respectively. According to \cite{3gpp.38.101}, the transmit power of UT is set as $25$ dBm. In addition, the carrier frequency $f_\textrm{c}$ is set as $28$ GHz, while $N_\textrm{sub}$ (the number of subcarriers) and $N_\textrm{cp}$ (the number of samples in the CP) are set as $128$ and $16$, respectively. Furthermore, in order to ensure good sensing performance, we assume that the subcarrier spacing $\Delta f$ is as large as $100$ KHz and  {consequently} the sampling interval is as small as $7.8125\times10^{-8}$s. As a result, the length of the OFDM symbol is $11.25\mu$s upon considering the length of the CP.  
		
		In each simulation trial, there are a total of 3 perceived targets, unless stated otherwise. Specifically, the velocities of the target vehicles are  randomly generated in the range of $[0, 40]$ m/s, and the distances between the target vehicles and the RRU are randomly generated in the range of $[20, 90]$ m. Moreover, DOAs of the received signals are uniformly generated from $[0,\pi]$. Additionally, the coordinates of the RRU are set as {$(0,0)$} m, while the UT coordinates are set as {$(80,0)$} m. Therefore, the perceived ranges between the UT and RRU can be computed as the sum of the distance of UT-targets and the distance of targets-RRU. 
		Additionally, static objects serving as reflectors are randomly set around the moving targets in each simulation. By adopting the cluster channel model, the number of static reflectors set around each moving target is randomly chosen to be between $2\sim 7$. These static objects are randomly generated within a circle with the target as its centre and a radius of $8$ meters. 
		For simplicity, all the target vehicles and static objects are modelled as point sources with a radar cross-section area of 1~m$^2$, as assumed in [13]. Moreover, the randomly set CFO is equivalent to the Doppler frequency shift induced by a velocity of $15\sim65$ m/s, while the randomly set TO is equivalent to the time delay induced by a range of $55\sim95$ meters.

\begin{figure}[tbp]
			\centering
			\includegraphics[width=3.5in]{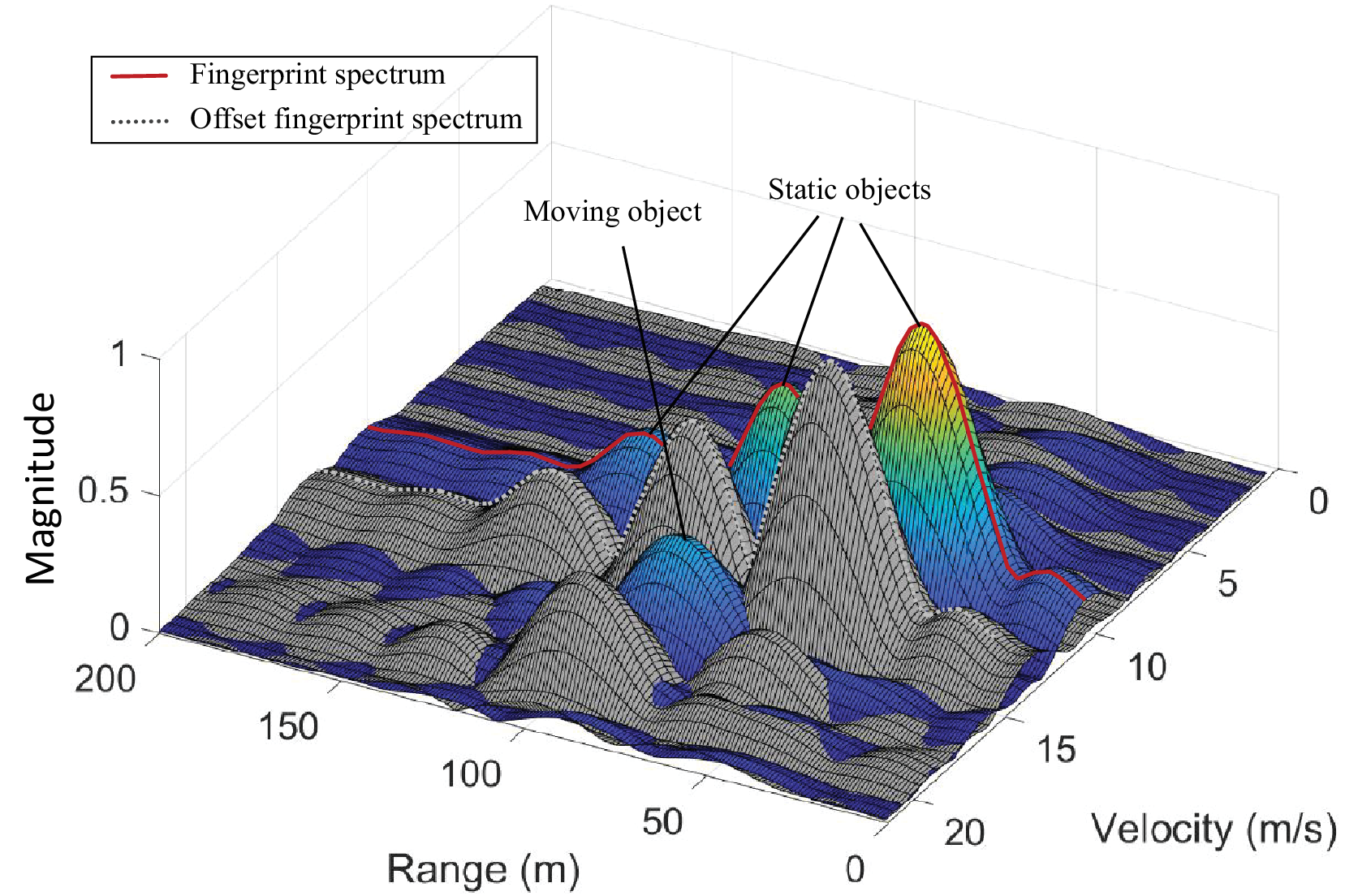}
			\caption{ {An implementation of  $\bar{\bf Y}$}. The red line represents the selected fingerprint spectrum.}
			\label{2D-FFTimage}
		\end{figure}

		In Fig. \ref{2D-FFTimage} we present a realization of $\bar{\bf Y}$ in a clutter-rich, and asynchronous PVN. To improve the resolution of sensing,  a $25$ times zero-padding and a $5$ times zero-padding are utilized in the range direction and the velocity direction, respectively. The blue 3D spectrum and the grey one represent the initial spectrum and the offset spectrum after CFO and TO drift. The set CFO and TO are equivalent to the Doppler frequency shift induced by a velocity of 3 m/s, and the time delay resulting from a distance of 10 m, respectively. The red outlines shown in both sub-figures are selected as the fingerprint spectrum sequence, while the dashed grey line represent the offset fingerprint spectrum sequence. As depicted in the figure, the offset fingerprint spectrum is the circular-shift fingerprint spectrum. Thus, we can estimate the CFO and TO by determining the number of shifts of the fingerprint spectrum.

		\begin{figure}[tbp]  
			\centering
			\includegraphics[width=3.5in]{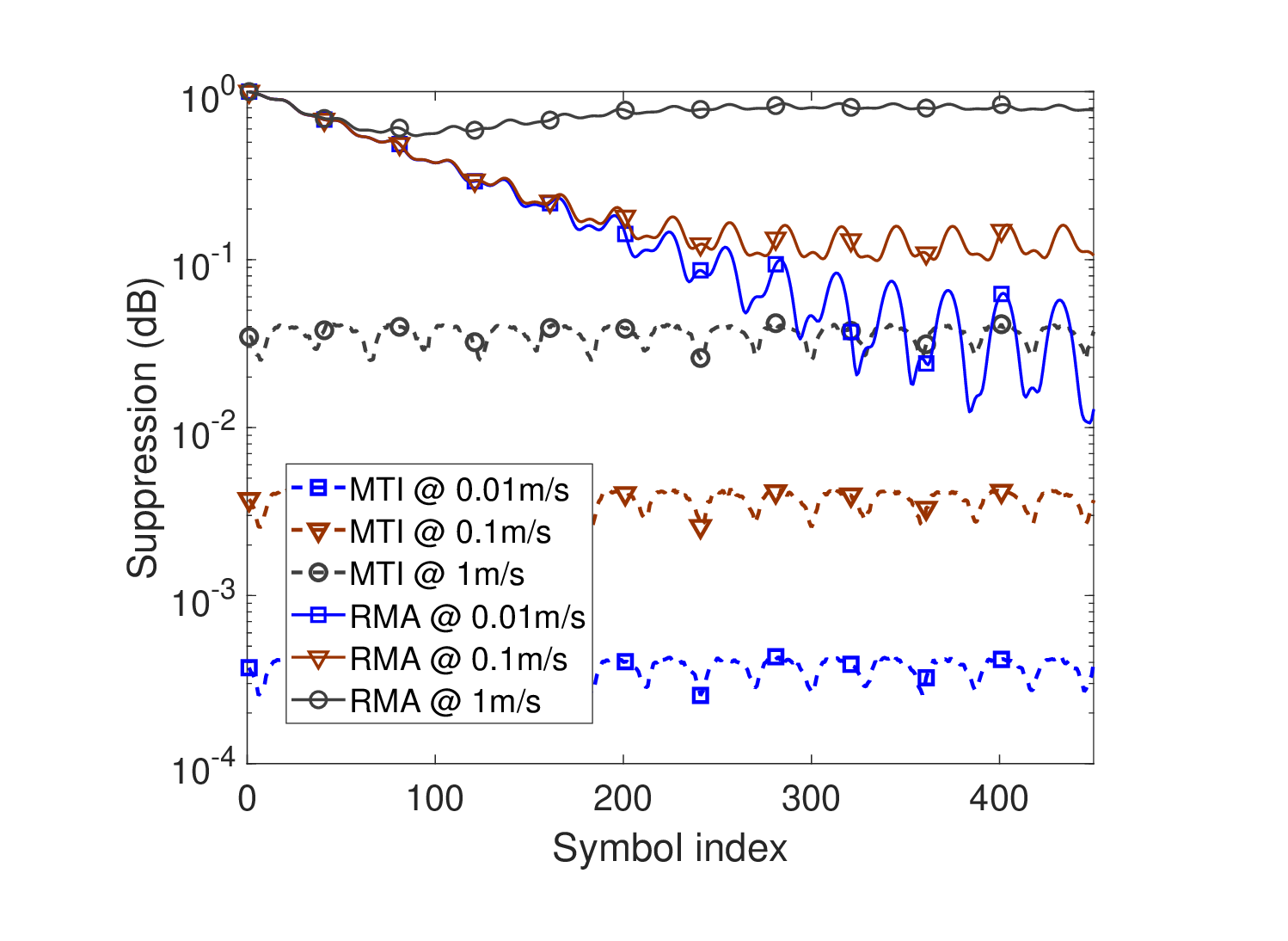}
			\caption{Clutter suppression performance of MTI and RMA based algorithms under  {different CFO. The specific velocity in the legend, such as $1$ m/s, represent that the set CFO is equivalent to the Doppler frequency shift induced by a velocity of $1$ m/s.}}
			\label{clutterRemovalComparing}
		\end{figure}
		
		\begin{figure}[tbp]	
			
			\begin{minipage}[t]{0.5\linewidth}
				\centering
				\subfigure[SNR vs MSE of ranging]{
					\includegraphics[width=3.5in]{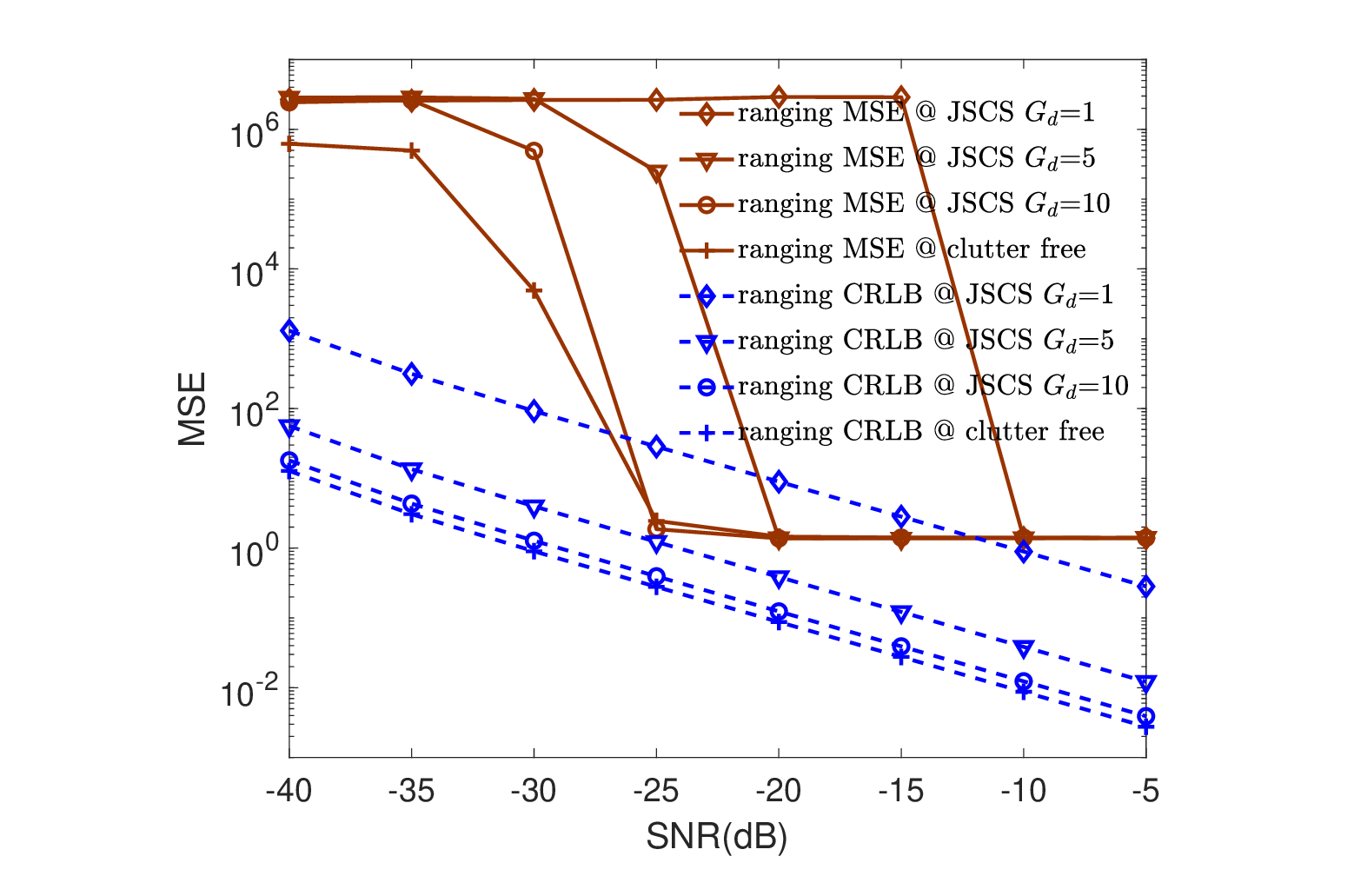}}
				
				
			\end{minipage}%
			
			\begin{minipage}[t]{0.5\linewidth}
				\centering
				\setcounter{subfigure}{1}\subfigure[SNR vs MSE of velocity estimation]{
					\includegraphics[width=3.5in]{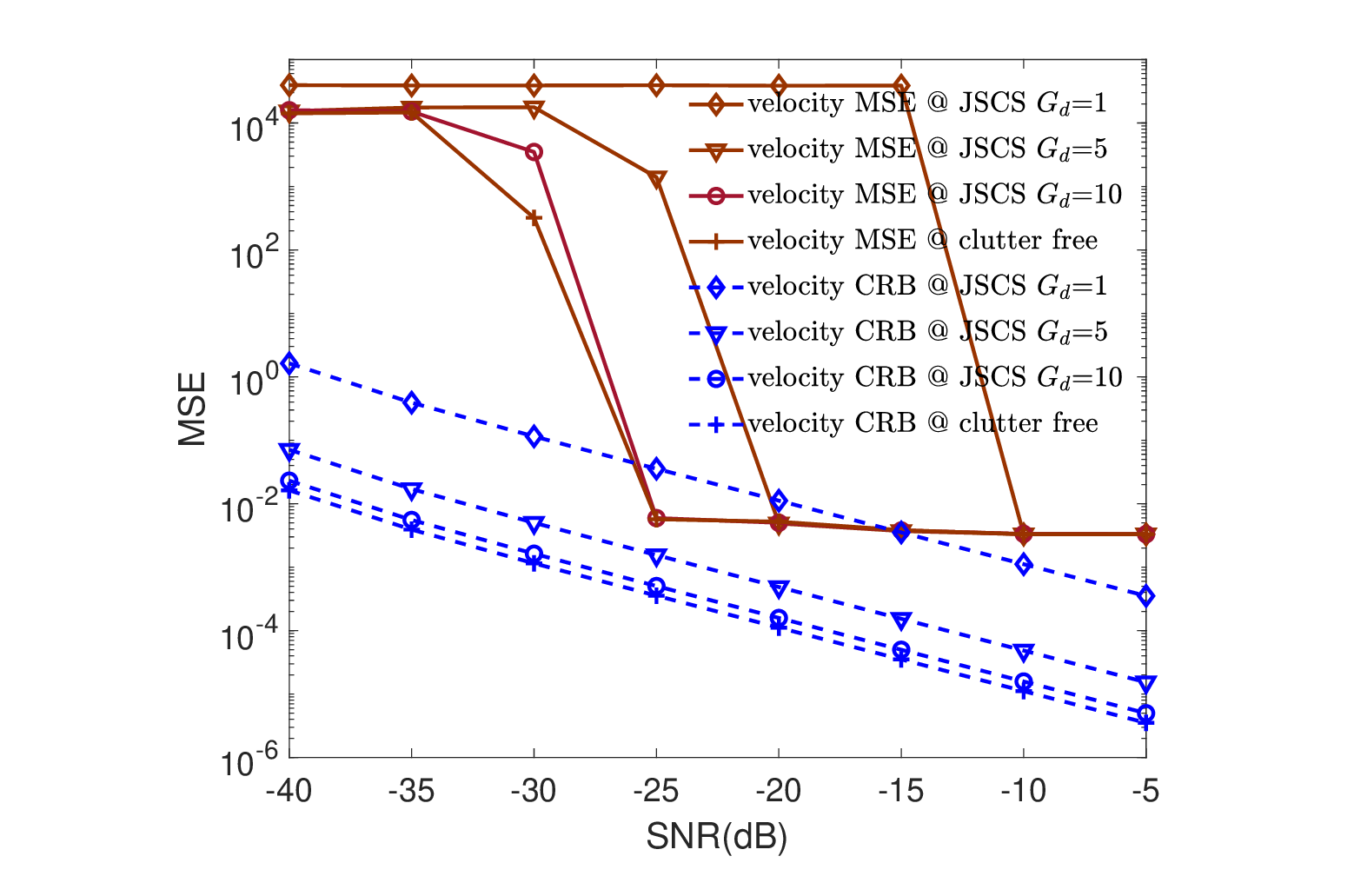}}
				
			\end{minipage}
			\caption{MSE and CRLB of CFO/TO vs SNR using MTI with $G_\textrm{d}=1, 5, 10$.}
			\label{CRLBrvcomparing}
		\end{figure}

		In Fig. \ref{clutterRemovalComparing}, we compare the performance of the RMA and the MTI clutter suppression algorithm with $G_\textrm{d} =1$. The clutter suppression performance  {in vertical axis is} $\rho_{\textrm b}/\rho_{\textrm a}$, where $\rho_{\textrm b}$ and $\rho_{\textrm a}$ represent the ratio of clutter power to  the power of the desired signal components before and after clutter suppression, respectively. It is observed that both the convergence rate and the clutter suppression performance of MTI is better than those of RMA. However, although MTI and RMA  is effective under existence of small CFO, the performance of both RMA and MTI degrades as the CFO increases. Therefore, to achieve better clutter suppression performance, the synchronization algorithm should be implemented before clutter suppression. Additionally, there are periodic fluctuations on each curve. The periodic fluctuations generate since both $\rho_{\textrm b}$ and $\rho_{\textrm a}$ are periodic functions with respect to the symbol index, and their periodic is determined by the velocities of the targets and CFO according to the definition of $\rho_{\textrm b}$ and $\rho_{\textrm a}$.

\begin{figure}[tbp]	
			\begin{minipage}[t]{0.5\linewidth}
				\centering
				\subfigure[SNR vs MSE of residual velocity ambiguity]{
					\includegraphics[width=3.5in]{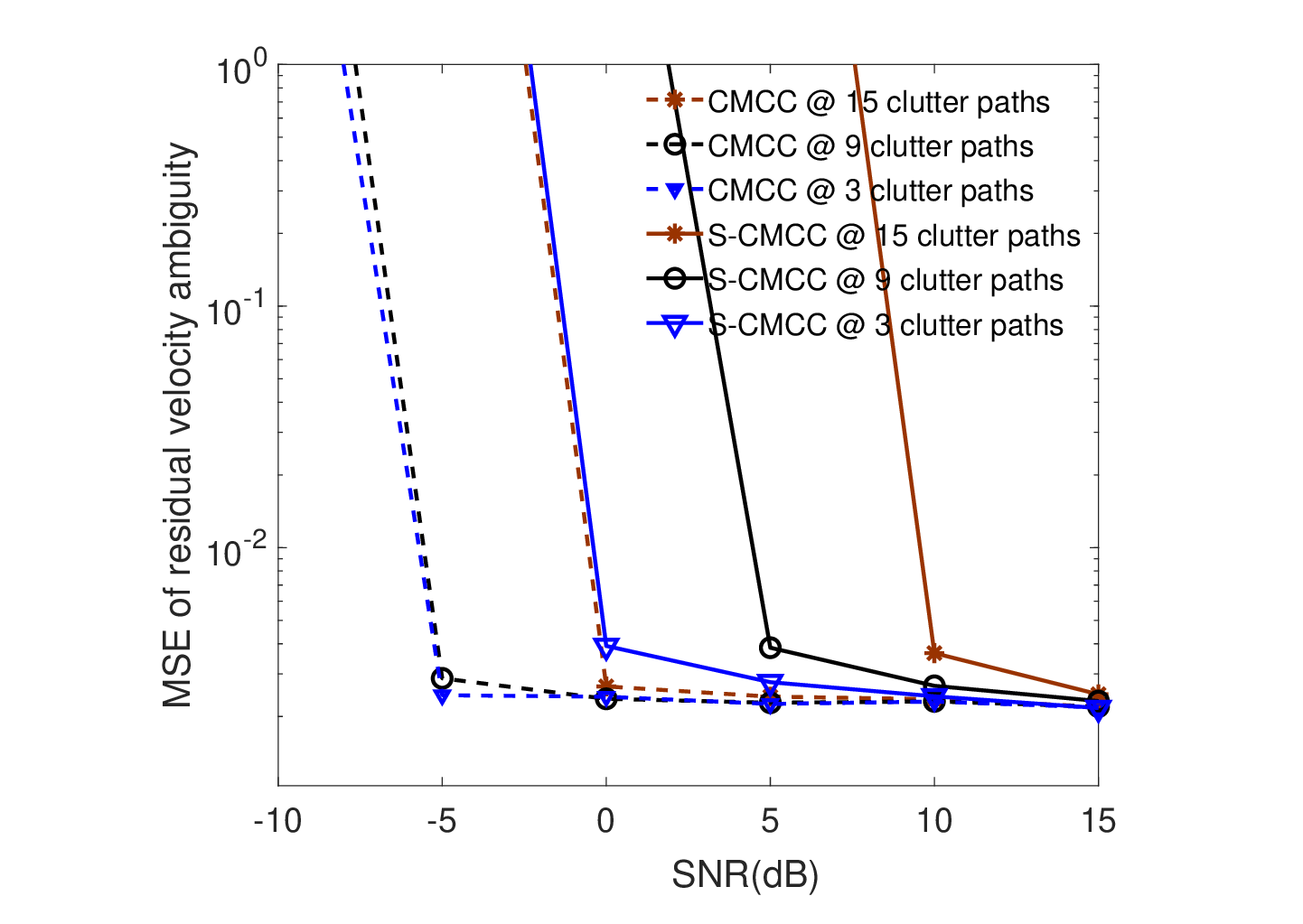}}		
			\end{minipage}%
			
			\begin{minipage}[t]{0.5\linewidth}
				\centering
				\setcounter{subfigure}{1}\subfigure[SNR vs MSE of residual range ambiguity]{
					\includegraphics[width=3.5in]{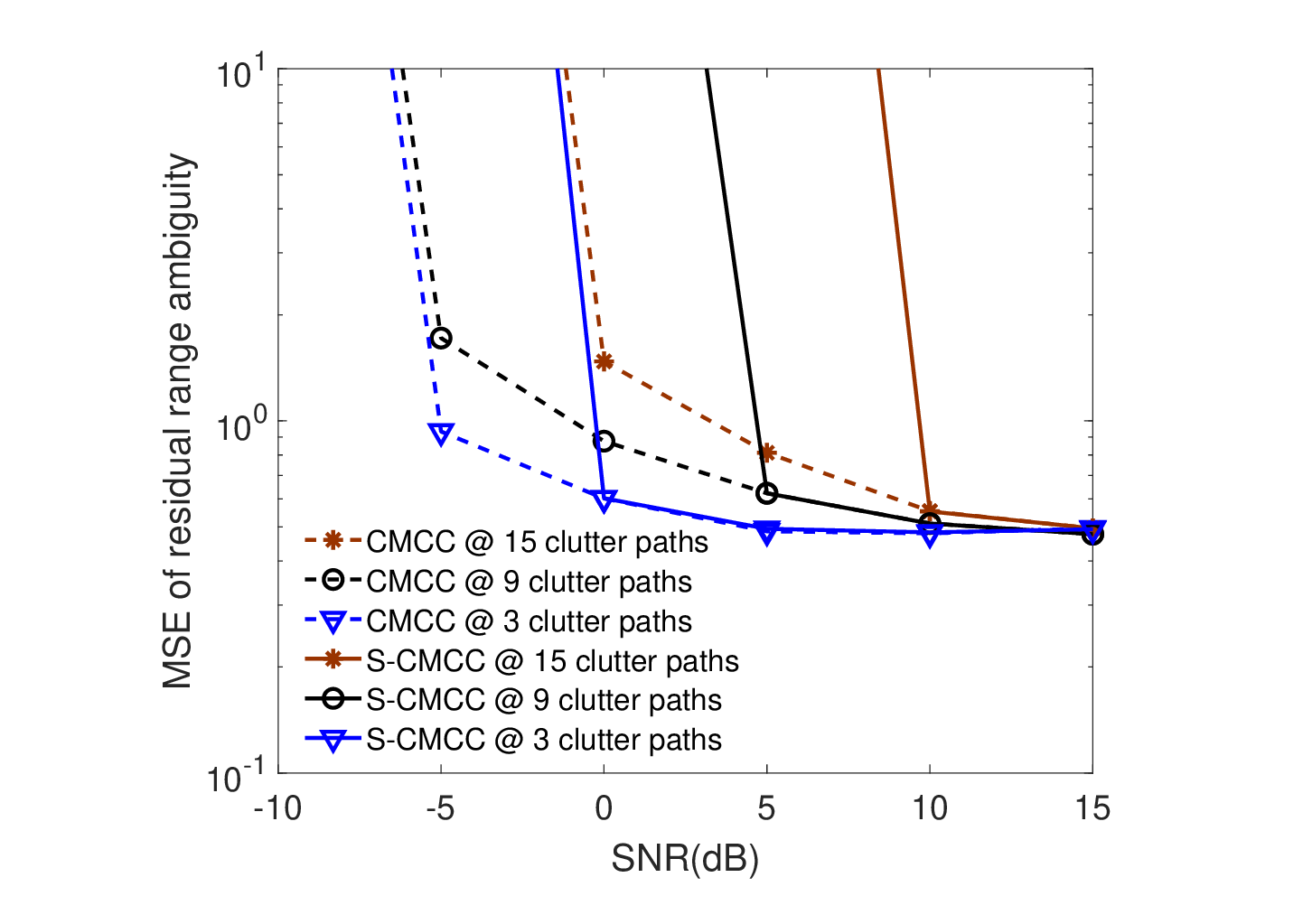}}	
			\end{minipage}
			\caption{MSE of residual velocity and range ambiguity vs SNR with \{3, 9, 15\} clutter paths and simplified/unsimplified estimation algorithm.}
			\label{MSErvHamming3-9-15clutter}
		\end{figure}

		In Fig. \ref{CRLBrvcomparing}, we evaluate the MSE performance and the CRLB of the proposed sensing scheme aided by MTI clutter suppression algorithm (denoted as ``JSCS" in the legend) against the  {clutter-free} range-velocity estimation  {without conducting MTI clutter suppression. Concretely,
			the MSE of the estimated velocity and the estimated range} are defined as
		${\textrm {MSE}_\textrm{V}} = \frac{1}{N_{\textrm {simu}}}\sum_{n=1}^{N_{\textrm {simu}}}[\frac{c\Delta f(\xi_{\textrm{o}}- {\hat\xi_{{\textrm o},n}})}{2f_\textrm{c}}]^2,
		$ and ${\textrm {MSE}_\textrm{R}} = \frac{1}{N_{\textrm {simu}}}\sum_{n=1}^{N_{\textrm {simu}}}[c(\tau_\textrm{o}- {\hat\tau_{{\textrm o},n}})]^2,$
		where $N_{\textrm {simu}}=1000$, $ {\hat\xi_{{\textrm o},n}}$, and $ {\hat\tau_{{\textrm o},n}}$ are the number of Monte Carlo simulations, the estimated $\xi_{\textrm{o}}$ in $n$th simulation, and the estimated $\tau_\textrm{o}$ in $n$th simulation, respectively. Specifically, Fig. \ref{CRLBrvcomparing} (a) and Fig. \ref{CRLBrvcomparing} (b) characterize the ranging estimation performance and the velocity estimation performance, respectively. As depicted in the figure, the MTI aided sensing has higher CRLB and MSE than that of the estimation in clutter-free environment. 
		However, with selected $G_\textrm{d}$, the CRLB of MTI-aided clutter suppression can be further improved to approach the CRLB corresponding to the clutter-free scenarios, which serves as the lower bound. Another notable phenomenon is the gap observed between the MSE and the CRB. There are three main reasons for this phenomenon: 1) when SNR falls below a certain threshold, the MSE degrades severely due to of the threshold phenomenon \cite{Optimum}; 2) conversely, when SNR surpasses a threshold, the MSE is primarily limited by the resolution of range/velocity in 2D-DFT rather than SNR; 3) the MSE of 2D-DFT-based is not exceptional. However, it can be improved by replacing it with more advanced estimation algorithms such as super-resolution estimation algorithms \cite{schmidt1986multiple,roy1989esprit}.
		In Fig. \ref{MSErvHamming3-9-15clutter}, we evaluate the synchronization performance of the proposed CMCC and S-CMCC algorithm in practical scenarios with different number of clutter path. 
		As presented in the figure, CMCC has much better performance than S-CMCC in all scenarios. This phenomenon emerges mainly since the 1-D ML CFO estimation in S-CMCC will definitely performs worse than the 2D ML estimation in CMCC, particularly when the SNR is small. Another obvious phenomenon is, the synchronization performance degrades when the path number of clutter increases for both CMCC and S-CMCC. The underlying reason for this phenomenon is that, at a constant SNR, the power distribution on the fingerprint spectrum tends to become uniform with an increasing path number of clutter (assuming the static objects reflecting clutter are randomly distributed). Then, a more uniform power distribution essentially diminishes the distinctiveness of a fingerprint spectrum, resulting in a reduction of the correlation peak in (\ref{unsimplified}) and (\ref{hatn}). This, in turn, reduces the MSE of synchronization. Therefore, both CMCC and S-CMCC cannot be applied in situations with a uniform-power-distribution fingerprint spectrum. To address this issue, transmitter beamforming can be utilized to select clutter reflectors and alter the shape of fingerprint spectrum. 
		Additionally, we have included Fig. 8 to assess the performance of the derived theoretical bound for CMCC. As depicted in the figure, the maximum difference between the theoretical bound and the MSE is approximately  0.4 m$^2$. 
		The difference illustrates the robustness of the theoretical bound.

		\begin{figure}[tbp]
			\centering
			\includegraphics[width=3.5in]{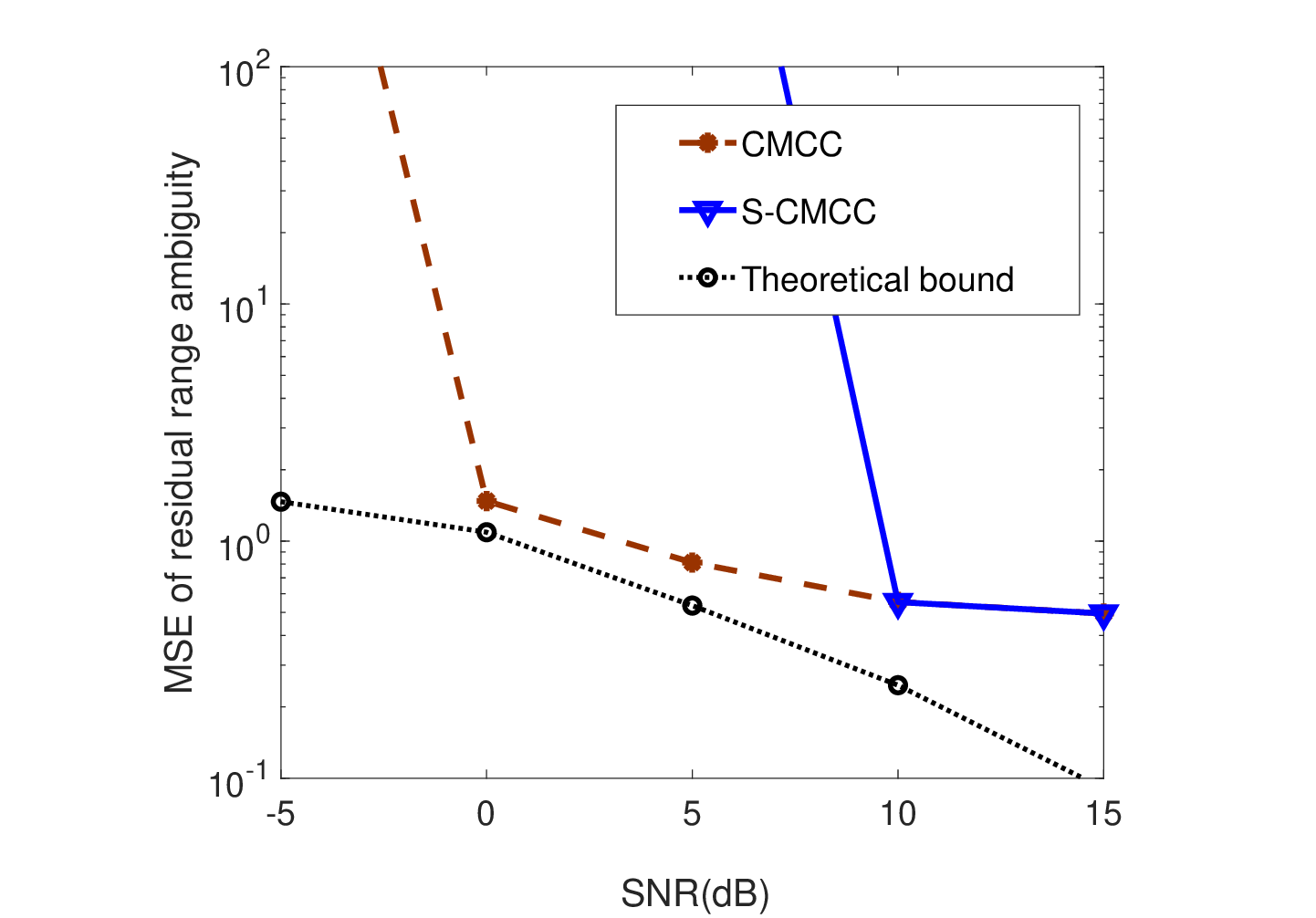}
			\caption{MSE and theoretical bound for residual range ambiguity vs. SNR.}
			\label{MSErwindow15clutter}
		\end{figure}

		\begin{figure}[tbp]
			\centering
			\includegraphics[width=3.5in]{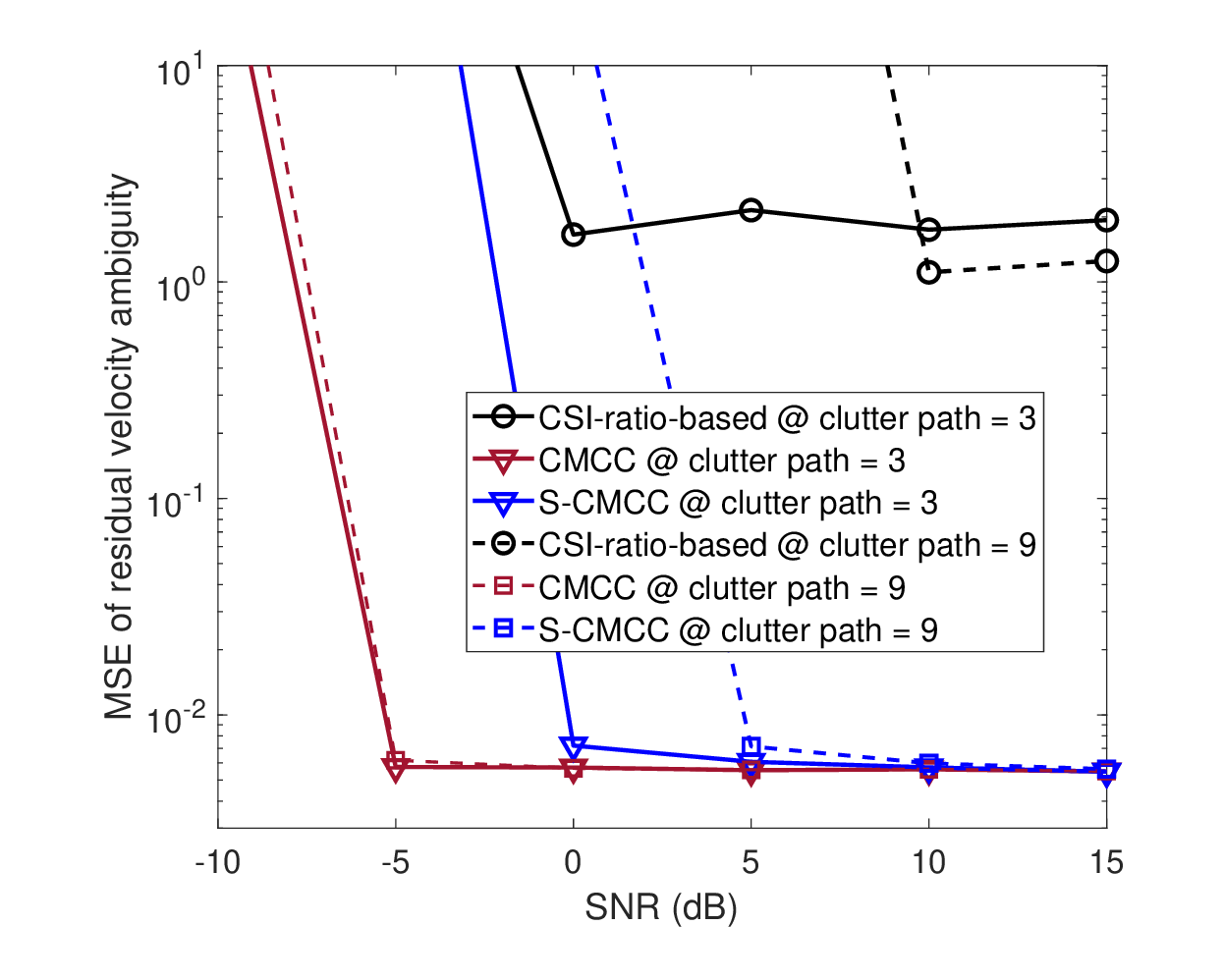}
			\caption{Comparison of MSE for velocity estimation in asynchronous systems.}
			\label{figure3}
		\end{figure}

		\begin{figure}[tbp]
			\centering
			\includegraphics[width=3.5in]{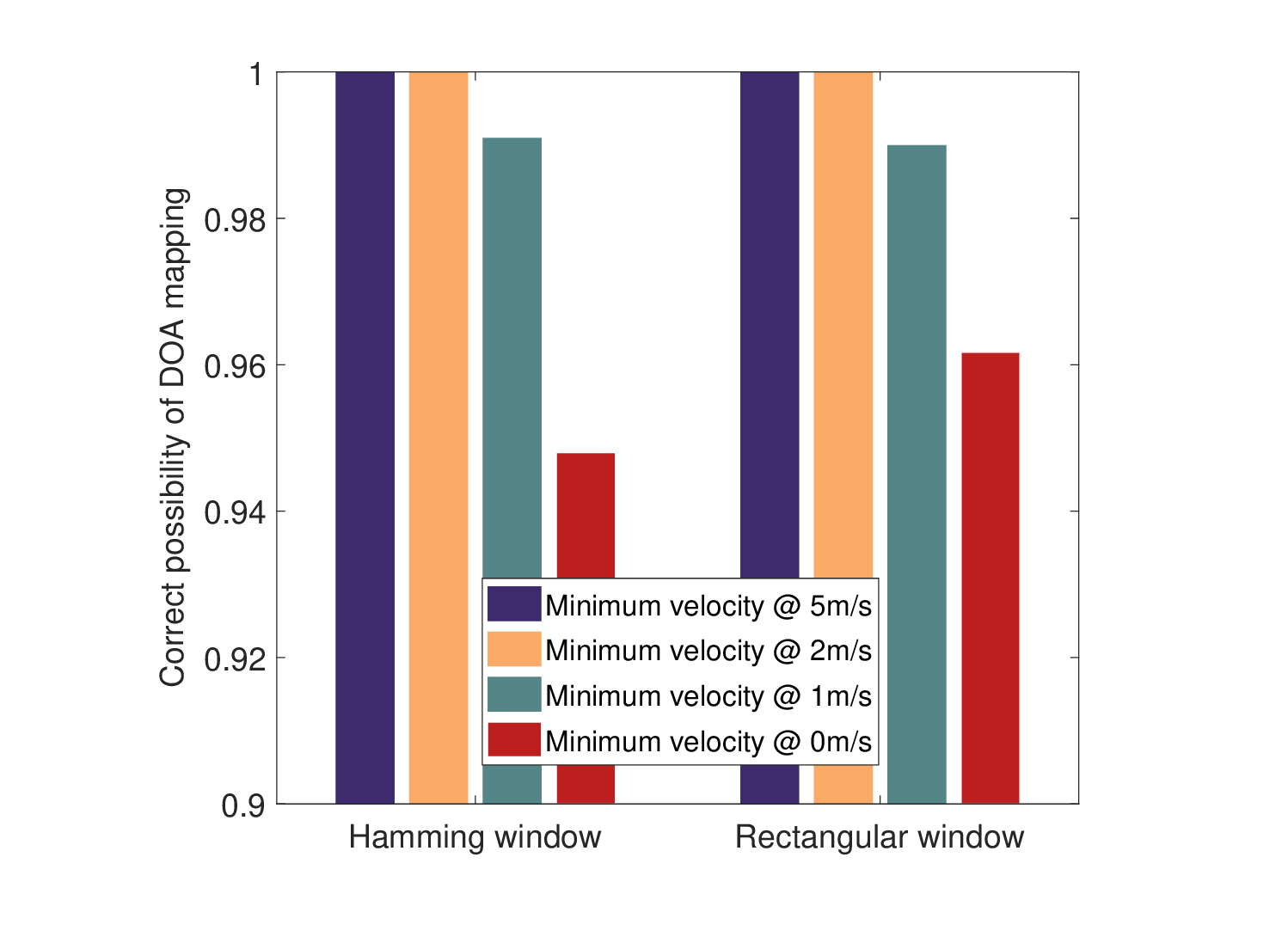}
			\caption{The probability of success of the proposed DOA association algorithm with Hamming and rectangular windows.}
			\label{CorrectPossibilityofDOAmapping}\label{DOAmapping}
		\end{figure}

		In Fig. \ref{figure3}, we evaluate the synchronization performance of the CMCC and CSI-ratio based schemes with different numbers of clutters. The simulation configurations are identical to those corresponding to Fig. 7. It is important to note that the CFO is not explicitly obtained in the state-of-the-art CSI-ratio based scheme, hence the MSE of the CFO estimate cannot be chosen as the evaluation metric.
		Specifically, the CSI-ratio based scheme does not directly estimate the CFO but rather reduces its impact on received signals by utilizing CSI ratio to estimate velocity. In contrast, the {proposed} CMCC algorithm directly estimates the CFO and TO, and then compensates for offsets existing in the estimated range and velocity. Therefore, we utilize the MSE of the estimated velocity under asynchronous systems to characterize synchronization performance.
		As observed in the figure, regardless of the number of clutters, our scheme consistently outperforms the CSI-ratio based scheme. However, both schemes exhibit degradation in performance with an increase in the number of clutters. The reason is that: given a certain SNR, with an increment of clutter paths, the power of signals reflected by targets will generally decrease, which results in worse estimation performance. Overall, the simulations demonstrate the superior performance of CMCC synchronization.

		In Fig. \ref{CorrectPossibilityofDOAmapping},  {the simplified algorithm as depicted by (\ref{newequ26}) is implemented in 1000 Monte Carlo simulations to evaluate the association performance. The minimum velocity in the legend represents the minimum difference among velocities of targets, which are randomly set in the simulation.} 
		In the simulation, the  {association performance corresponding to Hamming window and rectangular window are characterized for performance comparison. 
			As depicted in the figure, both the probability of success of Hamming and rectangular window equals 1 when the minimum velocity difference is set above $2$ m/s. Then, the performance deteriorates as the minimum velocity difference decreases. However, even the minimum velocity difference is set as $0$ m/s, both the probability of correct association corresponding to the Hamming and rectangular window is larger than 0.95. Moreover, as presented in the figure, the probability corresponding to rectangular window not worse than that of Hamming window at any minimum velocity difference.} This phenomenon can be interpreted as: the targets with similar velocities can be distinguished better with the window having narrower mainlobe. In other words, the window with narrower mainlobe have better resolution.  {As a result, to improve the performance, new spatial filter can also be designed.}
		
\begin{figure}[tbp]
			\centering
			\includegraphics[width=3.5in]{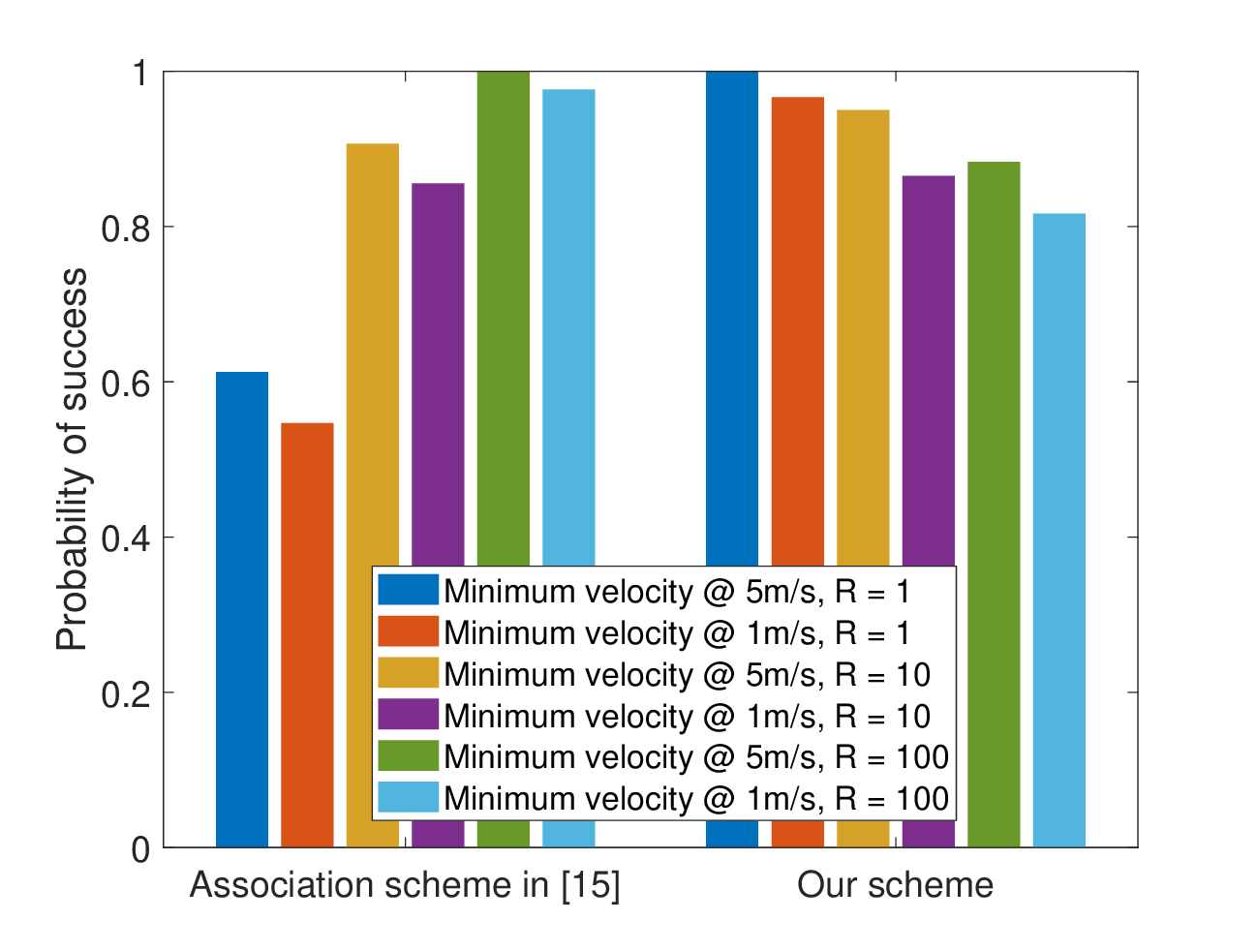}
			\caption{The probability of success of the proposed DOA association algorithm and the association method described in \cite{ni2021uplink}.}
			\label{figure2}
		\end{figure}

		To compare the performance of our DOA association scheme with that of the state-of-the-art solution of \cite{ni2021uplink}, we present their {probability of success} in Fig. \ref{figure2}. Since the benchmark association scheme only suits the scenarios that have LOS propagation paths, we conducted the comparison specifically in such scenarios. More specifically, to evaluate the performance of both algorithms in a comprehensive manner, we introduce two configurable parameters: the minimum difference among velocities of targets and the ratio (denoted as $R$) of the LOS-path power to the total power of NLOS paths. The former is adjusted to assess the effectiveness of the algorithms in resolving targets with similar velocities, while the latter is varied to assess the dependence on the signal strength of the LOS path relative to the NLOS paths.
		In our simulations, three objects with random velocities and locations are generated to ensure the generality of our setup. Each set of results in Fig. 11 is derived from 200 independent Monte Carlo experiments. As depicted in Fig. \ref{figure2}, for both our scheme and the scheme of \cite{ni2021uplink}, the probability of success slightly decreases as the minimum difference among velocities of targets is decreased. Another noteworthy observation is that the probability of success corresponding to the scheme of \cite{ni2021uplink} improves as $R$ becomes larger. {This happens since the establishment of association in \cite{ni2021uplink} is based on the assumption that the power of the LOS path is much greater than that of all the NLOS paths.} In contrast, the probability of success of our scheme deteriorates with the increase of $R$, {since the LOS path with significant power can introduce interference on the association metric calculated for NLOS paths during signal processing, and the impact increases with the power of the LOS path.} However, even in the scenario with the lowest probability of success, our scheme still outperforms the scheme of \cite{ni2021uplink}.

		\section{Conclusion}
		To suppress clutter and implement a practical range-velocity-DOA sensing scheme in asynchronous PVN, we introduce an MTI-clutter-suppression-aided uplink sensing scheme, a DOA association algorithm, and a synchronization algorithm. Subsequently, we analyse the suppression performance and the CRLB of the MTI-clutter-suppression-aided sensing scheme. Simulations demonstrate the high-performance clutter suppression of our scheme. Additionally, the proposed DOA association algorithm exhibits a high probability of correct association, with simulations confirming that this probability is influenced by the chosen window function. Furthermore, we introduce the first synchronization algorithm applicable to NLOS-PVN scenarios, namely CMCC, and its simplified version, S-CMCC. The synchronization performance of both CMCC and S-CMCC is dependent on certain environmental factors, particularly the number of clutter paths. Finally, we derive the CRLB of the MTI-clutter-suppression-aided sensing scheme, the theoretical MSE of synchronization.
		Future enhancements can be attained by investigating the correlation between window functions and synchronization performance, or by designing more effective spatial window functions for the DOA association algorithm.

		
		%



		\ifCLASSOPTIONcaptionsoff
		\newpage
		\fi

		\bibliographystyle{IEEEtran}
		\bibliography{IEEEabrv,reference.bib}
		
		\newpage
		
		\begin{IEEEbiography}[{\includegraphics[width=1in,height=1.25in,clip,keepaspectratio]{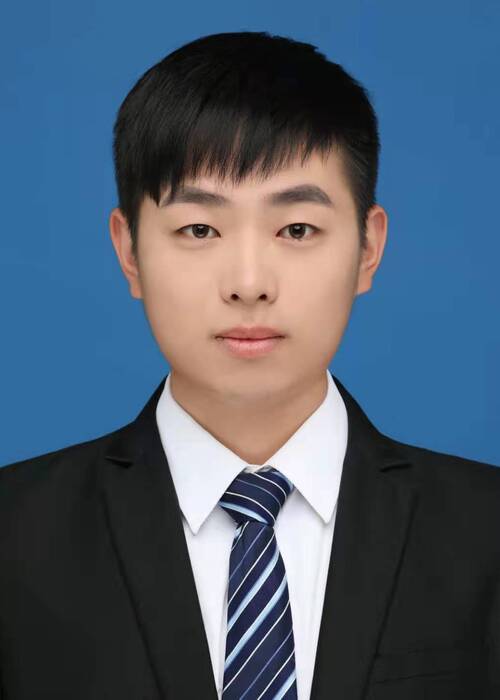}}]{Xiao-Yang Wang} received the B.S. degree in electronics and information science and technology from Beijing University of Posts and Telecommunications (BUPT), China, in 2020. He is currently pursuing the Ph.D. degree in information and communication engineering with the Key Laboratory of Universal Wireless Communications, Ministry of Education, BUPT. Since Sep. 2023, he has been a visiting Ph.D. student with the Department of Electronic and Electrical Engineering, University College London, UK. His current research interest includes integrated sensing and communications (ISAC), time/frequency synchronization, and signal processing in distributed massive MIMO systems.
		\end{IEEEbiography}
		%
		\begin{IEEEbiography}[{\includegraphics[width=1in,height=1.25in,clip,keepaspectratio]{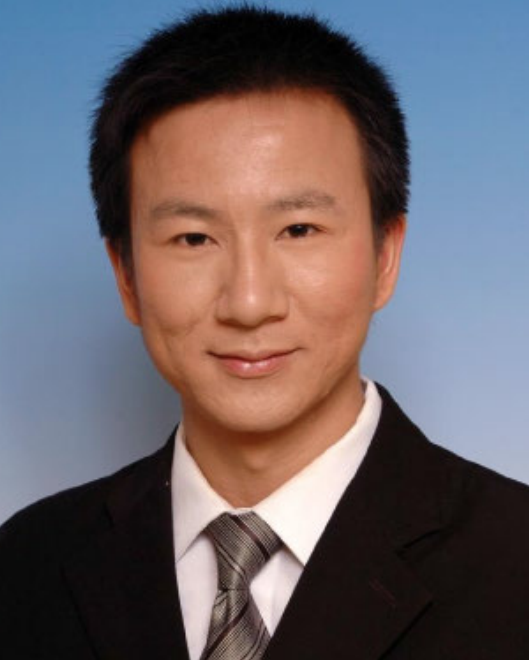}}]{Shaoshi Yang}
			(Senior Member, IEEE) 
			received the B.Eng. degree in information engineering from Beijing University of Posts and Telecommunications (BUPT), China, in 2006, and the Ph.D. degree in electronics and electrical engineering from University of Southampton, UK, in 2013. From 2008 to 2009, he was a Researcher with Intel Labs China. From 2013 to 2016, he was a Research Fellow with the School of Electronics and Computer Science, University of Southampton. From 2016 to 2018, he was a Principal Engineer with Huawei Technologies Co., Ltd., where he made significant contributions to Huawei’s products and solutions on 5G base stations, wideband IoT, and cloud gaming/VR. He was a Guest Researcher with the Isaac Newton Institute for Mathematical Sciences, University of Cambridge. He is currently a Full Professor with BUPT. His research interests include 5G/5G-A/6G, massive MIMO, mobile ad hoc networks, distributed artificial intelligence, and cloud gaming/VR. He is a standing committee member of the China Computer Federation (CCF) Technical Committee on Distributed Computing and Systems. He received numerous research awards from University of Southampton, Huawei, IEEE ComSoc, Xiaomi Foundation, China Association of Inventions, and China Industry-University-Research Institute Collaboration Association. He was/is an Editor of \textit{IEEE Systems Journal}, \textit{IEEE Wireless Communications Letters}, and \textit{Signal Processing} (Elsevier). For more details of his research progress, please refer to https://shaoshiyang.weebly.com/.
		\end{IEEEbiography}
		
		\begin{IEEEbiography}[{\includegraphics[width=1in,height=1.25in,clip,keepaspectratio]{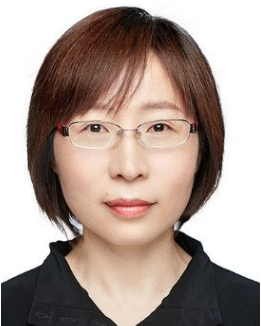}}]{Jianhua Zhang} (Senior Member, IEEE) received the Ph.D. degree in circuit and systems from Beijing University of Posts and Telecommunication (BUPT), China, in 2003. She is currently a Professor with BUPT, a Fellow of  China Institute of Communications, and the Director of the BUPT-CMCC Joint Research Center. She has 200+ technical papers published and 40+ patents granted. She received several paper awards, including 2019 SCIENCE China Information Hot Paper, 2016 China Comms Best Paper, and 2008 JCN Best Paper. She received several prizes for her contributions to ITU-R 4G channel model (ITU-R M.2135), 3GPP relay channel model (3GPP 36.814), and 3GPP 3D channel model (3GPP 36.873). She was also a member of 3GPP “5G channel model for bands up to 100 GHz.” From 2016 to 2017, she was the Drafting Group (DG) Chairperson of ITU-R IMT-2020 Channel Model and led the drafting of the ITU-R M. 2412 Channel Model Section. She is also the Chairwoman of the China IMT-2030 Tech Group—Channel Measurement and Modeling Subgroup and works on 6G channel model. Her current research interests include 5G-A and 6G wireless technologies, artificial intelligence, and data mining. She is an expert on channel modeling, channel emulator and OTA test for integrated sensing and communication, massive MIMO, millimeter wave, THz, and visible light communications. 
		\end{IEEEbiography}
		
		\begin{IEEEbiography}[{\includegraphics[width=1in,height=1.25in,clip,keepaspectratio]{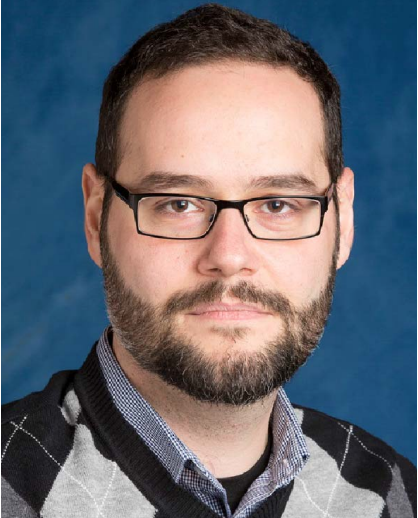}}]{Christos Masouros} (Fellow, IEEE) received the Diploma degree in electrical and computer engineering from  University of Patras, Greece, in 2004, and the M.Sc. by research and Ph.D. degrees in electrical and electronic engineering from The University of Manchester, U.K., in 2006 and 2009, respectively. In 2008, he was a Research Intern at Philips Research Labs, U.K. Between 2009 and 2010, he was a Research Associate with The University	of Manchester and a Research Fellow at Queen’s University Belfast between 2010 and 2012. In 2012, he joined University	College London as a Lecturer. He has held a Royal Academy of Engineering	Research Fellowship between 2011 and 2016. Since 2019, he has been a Full Professor of signal processing and wireless communications with the Information and Communication Engineering Research Group, Department of Electrical and Electronic Engineering, and affiliated with the Institute for Communications and Connected Systems, University College London.	His research interests lie in the field of wireless communications and signal processing with particular focus on green communications, large scale antenna systems, integrated sensing and communications (ISAC), interference mitigation techniques for MIMO, and multicarrier communications. He was the co-recipient of the 2021 IEEE SPS Young Author Best Paper Award, and the recipient of the Best Paper Awards in the IEEE GlobeCom	2015 and IEEE WCNC 2019 conferences. He has been recognized as an Exemplary Editor for IEEE Communications Letters and as an Exemplary Reviewer for  IEEE Transactions on Communications. He is a founding member and the Vice-Chair of the IEEE Emerging Technology Initiative on ISAC, the Vice Chair of the IEEE Special Interest Group on ISAC, and the Chair of the IEEE Special Interest Group on Energy Harvesting Communication Networks. He is an Editor of IEEE Transactions on Communications, IEEE Transactions on Wireless Communications, IEEE Open Journal of Signal Processing, and the Editor-at-Large of IEEE Open Journal of the Communications Society. He was an Associate Editor of IEEE Communications Letters and a Guest Editor for a number of special issues on IEEE Journal of Selected Topics in Signal Processing and IEEE Journal on Selected Areas in Communications.
		\end{IEEEbiography}

		\begin{IEEEbiography}[{\includegraphics[width=1in,height=1.25in,clip,keepaspectratio]{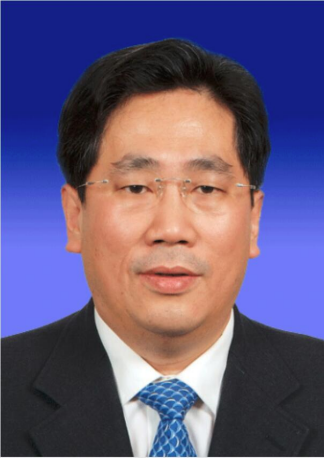}}]{Ping Zhang} (Fellow, IEEE) is a Professor with the School of Information and Communication Engineering, Beijing University of Posts and Telecommunications (BUPT), and an Academician of Chinese Academy of Engineering. He is the Director of the State Key Laboratory of Networking and Switching Technology, a member of IMT-2020 (5G) Experts Panel, and a member of Experts Panel for China’s 6G Development. He served as a Chief Scientist of National Basic Research Program (973 Program), an Expert in information technology division of National High-Tech Research and Development Program (863 Program), and a member of Consultant Committee on International Cooperation of National Natural Science Foundation of China. His research interests mainly focus on wireless communications.
		\end{IEEEbiography}
		

	\end{document}